\documentclass[11pt]{article}

\usepackage{amsmath}
\usepackage{amssymb}
\usepackage{graphicx}
\usepackage{color}
\usepackage{amsthm}
\usepackage{amsfonts}
\usepackage{amscd}
\usepackage{mathrsfs}
\usepackage{bm}
\usepackage{enumerate}
\usepackage{algorithmic, algorithm}
\DeclareMathOperator*{\argmin}{argmin}

\makeatletter

\usepackage{amsthm}\usepackage{amsfonts}\usepackage{amscd}\usepackage{mathrsfs}\usepackage{bm}\usepackage{enumerate}

\newcommand{\D}{\displaystyle}
\newcommand{\B}{\mathbf}

\newcommand{\dt}{\Delta\theta}

\newcommand{\myref}[1]{(\ref{#1})}
\newcommand{\mathd}{\mathrm{d}}

\newtheorem{remark}{\textbf{Remark}}[section]

\makeatother

\begin{document}

\title{Sparse Time-Frequency decomposition by dictionary learning}

\author{Thomas Y. Hou,\thanks{Applied and Comput. Math, MC 9-94, Caltech,
Pasadena, CA 91125. {\it Email: hou@cms.caltech.edu.}} \and
Zuoqiang Shi\thanks{Mathematical Sciences Center, Tsinghua University, Beijing, China, 100084.
{\it Email: zqshi@math.tsinghua.edu.cn.}} }

\maketitle

\begin{abstract}
In this paper, we propose a time-frequency analysis method to obtain instantaneous frequencies and the corresponding decomposition by solving an optimization problem.  In this optimization problem,
the basis to decompose the signal is not known {\it a priori}. Instead, it is adapted to the signal
and is determined as part of the optimization problem.
In this sense, this optimization problem can be seen as a dictionary learning problem.
This dictionary learning problem is solved by using the Augmented Lagrangian Multiplier method (ALM) iteratively.
 We further accelerate the convergence of the ALM method in each iteration by using the fast wavelet transform.
We apply our method to decompose several signals, including signals with poor scale separation, signals with outliers and polluted by noise and a real signal. The results show that this method can give accurate recovery
of both the instantaneous frequencies and the intrinsic mode functions.
\end{abstract}


\section{Introduction}

Nowadays we must process a massive amount of data in our daily life and the scientific research. Data analysis methods have played an important role in processing and analyzing these data. Most data analysis methods use pre-determined basis, including the most commonly used Fourier transform and wavelet transform. While these data analysis methods are very efficient in processing data, each component of the decomposition in general does not reveal the intrinsic physical information of these data due to the presence of harmonics in the decomposition. For example, application of these traditional data analysis methods to a modulated oscillatory chirp signal would produce many components. Thus, it is essential to develop a truly adaptive data analysis method that can extract hidden physical information such as trend and time varying cycles from these data and preserve the integrity of the physically meaningful components. To achieve this, we need to use a data-driven basis that is adapted to the signal instead of being determined {\it a priori}.

The EMD method \cite{Huang98,WH09} provides an efficient adaptive method to analyze nonlinear and nonstationary data. The EMD method is empirical in nature. Several methods with more clear mathematical structures have been proposed as a variant of the EMD method, see e.g.  synchrosqueezed wavelet transform \cite{DLW11},
Empirical wavelet transform \cite{DZ13}, variational mode decomposition \cite{Gilles13}.
Inspired by the EMD method and the recently developed compressed (compressive) sensing theory, we proposed
a data-driven time-frequency analysis method \cite{HS11,HS12}.
In this method, we formulate the problem as a nonlinear optimization problem and decompose the signal by looking for the sparsest representation of
a multiscale signal over a largest possible dictionary.

In our data-driven time-frequency analysis, we decompose a given signal $f(t)$ into the following form:
\begin{eqnarray}
\label{decomp-f}
  f(t)=\sum_{j=1}^M a_j(t)\cos\theta_j(t),\quad t\in \mathbb{R} ,
\end{eqnarray}
where $a_j(t),\; \theta_j(t)$ are smooth functions, $\theta_j'(t)>0,\;j=1,\cdots,M$
and $M$ is an integer that is given {\it a priori}. We assume that
$a_j(t)$ and $\theta_j'$ are less oscillatory than $\cos\theta_j(t)$.
We call $a_j(t)\cos\theta_j(t)$ as the Intrinsic Mode Functions (IMFs) and
$\theta'_j,\; j=1,\cdots, M$ the Instantaneous Frequencies \cite{Huang98}. The objective of our data-driven time-frequency analysis is to extract the Intrinsic Mode Functions and their Instantaneous Frequencies.

In \cite{HS12}, we proposed to decompose the signal by solving the following optimization problem:
 \begin{eqnarray}
\label{opt-l0}
\quad && \begin{array}{rcc}\vspace{-2mm}
&\mbox{Minimize} &M\\ \vspace{2mm}
&{\scriptstyle (a_k)_{1\le k\le M}, (\theta_k)_{1\le k\le M}}&\\
&\mbox{Subject to:}&\D f=\sum_{k=1}^M a_k\cos\theta_k,\quad\quad a_k\cos\theta_k\in \mathcal{D},
\end{array}
\end{eqnarray}
where $\mathcal{D}$ is the dictionary consist of all IMFs (see \cite{HS12} for its precise definition).
Further, an efficient algorithm based on matching pursuit and Fast Fourier transform has been proposed to
solve the above nonlinear optimization problem.
In a subsequent paper \cite{HST13}, we proved the convergence of our algorithm for periodic data
that satisfy certain scale separation property.

Basis pursuit is another powerful method to obtain the sparsest representation of a signal. Unforturnately, the basis pursuit cannot apply directly to \eqref{opt-l0},
since the dictionary $\mathcal{D}$ has infinitely many atoms. But if the number of IMFs, $M$, is known,
the idea of basis pursuit can be generalized to approximate the optimization problem \eqref{opt-l0}.
In this case, we can obtain the desirable decomposition by solving the following
dictionary learning problem:
 \begin{eqnarray}
  \label{opt-l1-intro}
  \min_{\B{x}, \theta_1\cdots, \theta_M}  \|\B{x}\|_1,\quad \mbox{subject to}\; \Phi_{\theta_1,\cdots,\theta_M}\B{x}=f,
\end{eqnarray}
where
\begin{eqnarray}
\Phi_{\theta_1,\cdots,\theta_M}=\left[ \Phi_{\theta_1},\cdots, \Phi_{\theta_M}\right] ,\label{fourier-theta-intro}
\end{eqnarray}
and $\Phi_{\theta_j},\; j=1,\cdots,M$ is the basis to decompose the signal $f$.
The specific form of the basis as a function of $\theta_j$ will be given in \myref{basis-cos}.

In \eqref{opt-l1-intro},
the basis $\Phi_{\theta_1,\cdots,\theta_M}$ is not known \textit{a priori}. It is determined by the phase functions
$\theta_j,\; j=1,\cdots,M$ and the phase functions are adaptive to the data.
We need to solve not only the optimal coefficients $\B{x}$ but also the optimal basis $\Phi_{\theta_1,\cdots,\theta_M}$.
In this sense, the problem described above can be seen as a
dictionary learning problem. Dictionary learning is a well-established subject in signal processing. Many efficient methods have been developed for this purpose \cite{AEB06,ESH07,LS00,MBPS09,SE10}. But the problem we study here has some important differences from a traditional dictionary learning problem.
In a traditional dictionary learning problem, a common setup starts with a training set,
a collection of training vectors. The atoms of the dictionary can be any
function. For the problem we consider, we only have one signal not a training set. Moreover, the atoms of the dictionary in our problem cannot be any function. If we do not put any constraint on the atoms of the dictionary, we may get only trivial decompositions. In order to get a reasonable result, the atoms of the dictionary are restricted to be nonlinear functionals of the phase function $\theta_j$.
At the same time, the phase functions are confined to be a low dimensional space to make sure that the overall degree of
freedoms is not too large. Then we can still get a reasonable decomposition with only one mesurement of the signal.

We need to develop a new method to solve this non-convex optimization problem, which is not required by a traditional dictionary learning problem. The key part is to find the phase functions. Once the phase functions
are known, we only need to solve a $l^1$ optimization problem to get the decomposition. Based on this observation, we
develop an iterative algorithm to solve \eqref{opt-l1-intro}. This algorithm starts from one initial guess of phase functions.
In each step, the phase functions are fixed and one $l^1$ optimization problem is solved. The phase functions will updated in the next iteration. This precedure is repeated until the iteration converges.
In each step, the Augmented Lagrangian Multiplier method (ALM) is used to solve the linear $l^1$ optimization problem.
We further accelerate the ALM method in each iteration by using the fast wavelet transform.
This method can be also generalized to decompose signals that contain outliers by enlarging the dictionary
to include impulses.

We will demonstrate the effectiveness of our method by applying it to decompose several multiscale data, include sythetic data and real data. For the data that we consider in this paper, we will demonstrate that we can recover both the instantaneous frequencies and the intrinsic mode functions very accurately. Even for those signals that are polluted by noise, we still can approximate the instantaneous frequencies and the IMFs with reasonable accuracy comparable to the noise level.

The remaining of the paper is organized as follows. In Section 2, we
give the formulation of our problem. In Section 3, an iterative algorithm
is introduced to solve the nonlinear optimization problem.
An accelerated procedure is introduced based on the fast wavelet transform
in Section 4. In Section 5, we generalize this method to deal with signals that
have outliers.  In Section 6, several numerical results are presented to demonstrate
the effectiveness of our method.
Finally, some concluding remarks are made in Section 7.

\section{Sparse time-frequency decomposition based on adaptive basis pursuit}

In this section, we will set up the framework of the sparse time-frequency decomposition. Let $\left\{V_l\right\}_{l\in \mathbb{Z}}$ be a multi-resolution approximation of $L^2(\mathbb{R})$ and
$\varphi$ the associated scaling function, $\psi$ the corresponding wavelet function. Assume that $\varphi$ is real and
$\widehat{\varphi}$ has compact support, $\mbox{supp}\left(\widehat{\varphi}\right)=(-s_\varphi,s_\varphi)$, where $\widehat{\varphi}$ is the
Fourier transform of $\varphi$ defined below,
\begin{eqnarray*}
  \widehat{\varphi}(k)=\frac{1}{2\pi}\int_\mathbb{R} \varphi(t)e^{-ikt}dt.
\end{eqnarray*}

For each $1\le j\le M$, we assume that $\theta_j'>0$. Since $\theta_j$
is a strictly monotonously increasing function, we can use $\theta_j$ as
a coordinate. Then we can define the following
wavelet basis in the $\theta_j$-coordinate,
\begin{equation}
\Pi_{\theta_j}=\left[ \psi_{l,n}\left(\theta_j\right)_{l, n\in \mathbb{Z},\atop 0 < l\le l_0},
\varphi_{l_0,n}\left(\theta_j\right)_{n\in \mathbb{Z}}
\right], \label{wavelet-envelope-j}
\end{equation}
where $l_0$ is a positive integer associated with the lowest frequency of the envelope. In practice, $l_0$ is usually determined by the time span of the signal. And
\begin{eqnarray}
\label{basis-env}
  &&\psi_{l,n}(\theta_j)=\frac{1}{\sqrt{2^{l}s_\varphi}}\psi\left(\frac{\theta_j}{2^{l}s_\varphi}-n\right),\\
&&\varphi_{l,n}(\theta_j)=\frac{1}{\sqrt{2^{l}s_\varphi}}\varphi\left(\frac{\theta_j}{2^{l}s_\varphi}-n\right) .
\end{eqnarray}
In this paper, we use the Meyer wavelet to construct the basis.

We assume that the envelope is sufficiently smooth, $a_j\in C^\infty$.
Thus, $a_j$ has a sparse representation over the wavelet basis
$\Pi_{\theta_j}$, and the corresponding IMF, defined as $a_j\cos\theta_j$,
has a sparse representation over the following basis:
\begin{eqnarray}
&&\hspace{-2mm}\Phi_{\theta_j}=(\cos\theta_j)\Pi_{\theta_j}\nonumber\\
&=&
\left[ \left(\psi_{l,n}(\theta_j)\cos\theta_j\right)_{l, n\in \mathbb{Z},\atop 0< l\le l_0},\left(\varphi_{l_0,n}(\theta_j)\cos\theta_j\right)_{n\in \mathbb{Z}}
\right]\quad\label{basis-cos} .
\end{eqnarray}
This in turn implies that the signal $f(t)$ given
in \myref{decomp-f} would have a sparse representation over the
combination of all $\Phi_{\theta_j},\; j=1,\cdots,M$:
\begin{eqnarray}
\Phi_{\theta_1,\cdots,\theta_M}=\left[ \Phi_{\theta_1},\cdots, \Phi_{\theta_M}\right] .\label{fourier-theta}
\end{eqnarray}
Based on this observation, we propose to decompose the signal $f(t)$ in \myref{decomp-f} by looking for the sparsest representation over $\Phi_{\theta_1,\cdots,\theta_M}$:
\begin{eqnarray}
  \label{opt-l0}
  \min_{\B{x}, \theta_1\cdots, \theta_M}  \|\B{x}\|_0,\quad \mbox{subject to}\; \Phi_{\theta_1,\cdots,\theta_M}\B{x}=f ,
\end{eqnarray}
where $\B{x}=\left[\B{x}_1,\cdots,\B{x}_M\right]^T$ and $\B{x}_j,\; j=1,\cdots, M$ are the corresponding coefficients of $a_j\cos\theta_j$ when we project $a_j\cos\theta_j$ over the basis $\Phi_{\theta_j}$.

Thanks to the development of the compressive sensing, we know that the
minimization using the $l^0$ norm can be relaxed to the convex minimization using the $l^1$ norm in many cases. We also
apply this relaxation to simplify the optimization problem. This gives rise to the
following nonlinear $l^1$ optimization problem:
 \begin{eqnarray}
  \label{opt-l1}
  \min_{\B{x}, \theta_1\cdots, \theta_M}  \|\B{x}\|_1,\quad \mbox{subject to}\; \Phi_{\theta_1,\cdots,\theta_M}\B{x}=f .
\end{eqnarray}
This nonlinear optimization problem is actually a dictionary learning problem. In this problem,
not only the coefficients are unknown, but the basis over which the signal has a sparse representation is also unknown {\it a priori}. The basis is determined by the phase functions $\theta_1,\cdots,\theta_M$. Both the coefficients $\B{x}$ and the phase functions need to be determined in the optimization process.

\begin{remark}
  In this paper, we construct the basis by utilizing the wavelet basis instead of the overcomplete Fourier basis which was used in our previous paper \cite{HS12}.
We choose the wavelet basis over the overcomplete Fourier basis because there are fast decomposition and reconstruction algorithms using the wavelet basis. This feature makes our algorithm very
efficient. For periodic signals, we can use the standard Fourier basis, which can be made very efficient by using the Fast Fourier transform. But
the Fourier basis works only for periodic signals. For general nonperiodic signals, the wavelet basis seems to be more robust although it still suffers from the "end effect" near the boundary of the signal.
\end{remark}

\section{Algorithm based on the Augmented Lagrangian Multiplier Method}

Inspired by the algorithm in \cite{HS12}, we propose the following iterative algorithm to solve the nonlinear $l^1$ optimization problem \myref{opt-l1}.
\begin{algorithm}[H]
\floatname{algorithm}{Algorithm}
\caption{(Gauss-Newton type iteration)}
\label{alg:guass-newton}
\begin{algorithmic}[1]
\REQUIRE Initial guess of phase functions $\theta_j^0,\; j=1,\cdots,M$, $\eta=l_0$, where $l_0$ is same as that in \myref{wavelet-envelope-j}.
\ENSURE Phase functions and corresponding envelopes: $\theta_j,\;a_j,\quad j=1,\cdots,M$.
\WHILE{$\eta\ge 1$}
\WHILE{ $\D \sum_{j=1}^M\|\theta_j^{n+1}-\theta_j^{n}\|_2>\epsilon_0$}
\STATE Solve the following $l^1$ optimization problem:
\begin{eqnarray}
\label{opt-mul-linear}
&&\hspace{-1cm}\left(\B{\widetilde{a}}^{n+1},\B{\widetilde{b}}^{n+1}\right)= \D \argmin_{\B{x},\B{y}}  (\|\B{x}\|_1+\|\B{y}\|_1), \\
&&\hspace{-1cm}\mbox{subject to}\quad \Phi_{\theta_1^n,\cdots,\theta_M^n}\cdot\B{x}
+\Psi_{\theta_1^n,\cdots,\theta_M^n}\cdot\B{y}=f,\nonumber
\end{eqnarray}
where $\Psi_{\theta_1^n,\cdots,\theta_M^n}=[\Psi_{\theta_1^n},\cdots,\Psi_{\theta_M^n}]$ and
\begin{eqnarray}
\Psi_{\theta_j^n}=(\sin\theta_j)\Pi_{\theta_j}
,\quad j=1,\cdots,M.\label{basis-sin}
\end{eqnarray}
\STATE Calculate the envelopes $a_j,\;b_j,\;j=1,\cdots M$:
\begin{eqnarray}
\label{eq:envs}
a_{j}=\Pi_{\theta^n_j}\cdot \B{\widetilde{a}}_j^{n+1},\quad
b_{j}=\Pi_{\theta^n_j}\cdot \B{\widetilde{b}}_j^{n+1},\quad
\end{eqnarray}
\STATE  Update $\theta^n_j,\; j=1,\cdots,M$:
\begin{eqnarray*}
\dt_j' &=& P_{V_\eta(\theta^n_j)}\left(\frac{d}{dt}\left(\arctan \left(\frac{b_j}{a_j}\right)\right)\right),\\
\dt_j&=&\int_0^t\dt_j'(s)ds,\quad
\theta^{n+1}_j=\theta^n_j-\beta_j \dt_j,
\end{eqnarray*}
where $\beta_j\in [0,1]$ is chosen to make sure that
 $\theta_j^{n+1}$ is monotonically increasing:
\begin{eqnarray}
\beta_j=\max\left\{\alpha\in [0,1]: \frac{d}{dt}\left(\theta_j^n-\alpha
\dt_j\right)\ge 0\right\}.\nonumber
\end{eqnarray}
Here $P_{V_\eta(\theta^n_j)}$ is the projection operator to the space $V_\eta(\theta^n_j)$ and
\begin{eqnarray}
V_\eta(\theta)=\mbox{span}\left[ \psi_{l,n}\left(\theta\right)_{l, n\in \mathbb{Z},
\atop \eta < l\le l_0},
\varphi_{l_0,n}\left(\theta\right)_{n\in \mathbb{Z}}
\right].\nonumber
\end{eqnarray}
\ENDWHILE
\STATE $\eta=\eta-1$.
\ENDWHILE
\end{algorithmic}
\end{algorithm}
This algorithm is essentially based on the Gauss-Newton iteration
which is derived by performing a local linearization around the phase functions in the current step. Both the matrix
$\Psi_{\theta_1^n,\cdots,\theta_M^n}$ in \myref{opt-mul-linear} and $b_j$ in \myref{eq:envs} come from this linearization
procedure. The coefficients, $b_j$, are just auxiliary functions in the iteration, and will tend to zero as the iteration converges.

In some sense, the above algorithm shares some common idea with that of  the K-SVD algorithm \cite{AEB06}. The step to solve the $l^1$
optimization problem is similar to the sparse coding stage in the K-SVD method and the step to update the phase functions is similar to the
codebook update stage. The main difference is in the codebook update stage. In our problem, the atoms of the dictionary
are not arbitrary function. It is parameterized by the phase functions. For this reason, we cannot use SVD to update the dictionary directly.
Instead, we use the Gauss-Newton algorithm to update the phase functions, which in turn updates the dictionary accordingly.

In the step to update the phase functions, we choose to update the instantaneous frequency (derivative of the phase
function) instead of updating the phase function directly. This is because
$\arctan \left(\frac{b_j}{a_j}\right)$ could be discontinuous if we do not add or subtract $2\pi$ at some particular points.
The choice of $\beta_j$ is to make sure that the new instantaneous frequency is positive.

The projection operator $P_{V_\eta(\theta_j^n)}$ is used to abate the
dependence on the initial guess.  The value of $\eta$ is gradually increased to improve the accuracy of our approximation to the
phase function so that it converges to the correct value. When
$\eta$ is small, $\Delta \theta'$ is confined to a small space.
In this small space, the objective functional has fewer extrema.
The iteration may find an good approximation for $\Delta \theta'$.
By gradually increasing $\eta$, we enlarge the space for $\Delta \theta'$
which allows for the small scale structures of $\Delta \theta'$
to develop gradually during the iterations.

In the numerical examples presented in Section \ref{sec:numerics}, we obtain the initial guess using
a simple approach based on the Fourier Transform. More precisely, we obtain the initial guess by
estimating the wavenumber by which the high frequency components are
centered around.

In the above algorithm, the most expensive part of the algorithm is to solve the $l^1$ optimization problem \myref{opt-mul-linear}. In this paper, we use an Augmented
Lagrange Multiplier (ALM) algorithm \cite{Bert82} to solve \myref{opt-mul-linear}. In order to simplify the notations, we denote
\begin{eqnarray}
\label{def-Theta}
  \Theta_{\theta_1^n,\cdots,\theta_M^n}=\left[\Theta_{\theta_1^n},\cdots,\Theta_{\theta_M^n}\right],
\end{eqnarray}
where
\begin{eqnarray*}
\Theta_{\theta_j^n}=\left[\Phi_{\theta_j^n},\Psi_{\theta_j^n}\right],\quad
j=1,\cdots,M.
\end{eqnarray*}

The ALM method operates on the augmented Lagrangian
\begin{eqnarray}
L(\B{p},\B{q})&=&\|\B{p}\|_1+\left<\B{q},f-\Theta_{\theta_1,\cdots,\theta_M}\B{p}\right>\nonumber\\
&&+\frac{\mu}{2}\|f-\Theta_{\theta_1,\cdots,\theta_M}\B{p}\|_2^2 .
\end{eqnarray}
A generic Lagrange multiplier algorithm \cite{Bert82} would solve \myref{opt-mul-linear} by repeatedly setting $\B{p}^{k+1}= \argmin_{\B{p}}L(\B{p},\B{q}^k)$, and
 then updating the Lagrange multiplier via $\B{q}^{k+1}=\B{q}^k+\mu (f-\Theta_{\theta_1,\cdots,\theta_M}\B{p}^k)$.

In this iteration, solving
$\min_{\B{p}}L(\B{p},\B{q}^k)$ is also very time consuming. Note that the matrix $\Theta_{\theta_1,\cdots,\theta_M}=\left[ \Theta_{\theta_1},\cdots, \Theta_{\theta_M}\right]$
is the combination of $M$ matrices with smaller size. It is natural to use the following sweeping algorithm to solve $\min_{\B{p}}L(\B{p},\B{q}^k)$ iteratively:
\begin{algorithm}[H]
\floatname{algorithm}{Algorithm}
\caption{(Sweeping)}
\label{alg:sweep}
\begin{algorithmic}[1]
\REQUIRE $\B{p}_j^0=0,\; j=1,\cdots,M$, $\mu>0$.
\WHILE{not converge}
\FOR{j=1:M}
\STATE Compute
$\B{r}_j^m=f-\sum_{l=1}^{j-1}\Theta_{\theta_l}\B{p}_l^{m+1}-\sum_{l=j+1}^{M}\Theta_{\theta_l}\B{p}_l^{m}$.
\STATE Compute $\D \B{p}_j^{m+1}=\argmin_{\B{p}_j} \|\B{p}_j\|_1+\frac{\mu}{2}\|\B{r}_j^m+\B{q}^k/\mu-\Theta_{\theta_j}\B{p}_j\|_2^2$.
\ENDFOR
\ENDWHILE
\end{algorithmic}
\end{algorithm}

Theoretically, we need to run the above sweeping process several times until the solution converges, but in practical computations, in order to save the
computational cost, we only run the sweeping process once. Combining this idea with the augmented Lagrange multiplier method, we obtain the following algorithm
to solve the $l^1$ optimization problem \myref{opt-mul-linear}:
\begin{algorithm}[H]
\floatname{algorithm}{Algorithm}
\caption{(Sweeping ALM)}
\label{alg:sweep-ALM}
\begin{algorithmic}[1]
\REQUIRE $\B{p}_j^0=0,\; j=1,\cdots,M$, $\B{q}^0=0$, $\mu>0$
\WHILE{not converge}
\FOR{j=1:M}
\STATE Compute
$\B{r}_j^k=f-\sum_{l=1}^{j-1}\Theta_{\theta_l}\B{p}_l^{k+1}-\sum_{l=j+1}^{M}\Theta_{\theta_l}\B{p}_l^{k}$.
\STATE Compute $\D \B{p}_j^{k+1}=\arg\min_{\B{p}_j} \|\B{p}_j\|_1+\frac{\mu}{2}\|\B{r}_j^k+\B{q}^k/\mu-\Theta_{\theta_j}\B{p}_j\|_2^2$.
\ENDFOR
\STATE $\B{q}^{k+1}=\B{q}^k+\mu \left(f-\sum_{j=1}^M \Theta_{\theta_j}\B{p}_j^{k+1}\right)$.
\ENDWHILE
\end{algorithmic}
\end{algorithm}

\section{A fast algorithm based on the discrete wavelet transform}
In this section, we propose an approximate solver to accelerate the computation of
$\D \B{p}_j^{k+1}=\arg\min_{\B{p}_j} \|\B{p}_j\|_1+\frac{\mu}{2}\|\B{r}_j^k+\B{q}^k/\mu-\Theta_{\theta_j}\B{p}_j\|_2^2$ which is the most expensive step in
Algorithm 1.

In this paper, we only consider the signal which is well resolved by the samples and the samples are uniformly distributed in time and the total number
of the samples is $N$. Based on these assumptions, we can approximate the continuous integral by the discrete summation and the integration error is
negligible. This greatly simplifies our calculations.

Now, we turn to simplify the optimization problem.
First, we replace the standard $L^2$ norm by a weighted $L^2$ norm which gives the following approximation:
\begin{eqnarray}
\label{weight-l1}
    \B{p}_j^{k+1}=\arg\min_{\B{p}_j} \|\B{p}_j\|_1+\frac{\mu}{2}\|\B{r}_j^k+\B{q}^k/\mu-\Theta_{\theta_j}\B{p}_j\|_{2,\theta_j}^2,
\end{eqnarray}
where $\|\B{g}\|_{2,\theta_j}^2=\sum_i g_i^2 \theta_j'(t_i)$.

Using the fact that $\mbox{supp}(\widehat{\varphi})=(-s_\varphi,s_\varphi)$, it is easy to check that
the columns of the matrix $\Theta_{\theta}$ are orthonormal under the weighted discrete inner product
\begin{eqnarray}
\label{weighted-inner-product}
\left<\B{g},\B{h}\right>_{\theta}
=\sum_{i=1}^N g_ih_i\theta'(t_i) =\B{g}\cdot(\theta'\B{h}),
\end{eqnarray}
where $\B{g}=\left[g_i\right],\; \B{h}=\left[h_i\right],\; \theta'\B{h}=\left[\theta'(t_i)h_i\right],\; i=1,\cdots,N$.

Using this property, it is easy to derive the following equality:
\begin{eqnarray}
  \label{eq:parsaval}
 \|\Theta_{\theta}\cdot \B{x}\|_{2,\theta}=\|\B{x}\|_2 .
\end{eqnarray}
We can also define the projection operator $P_{V(\theta)}$ to $V(\theta)$ space. Here
$V(\theta)$ is the linear space spanned by the columns of the matrix $\Theta_{\theta}$ and $P_{V(\theta)}$ is the projection operator to $V(\theta)$.
Since the columns of the matrix $\Theta_{\theta}$ are orthonormal under the weighted inner product \myref{weighted-inner-product}, projection
$P_{V(\theta)}$ can be calculated as follows:
\begin{eqnarray}
  P_{V(\theta)}(\B{r})=\Theta_{\theta}\cdot \widehat{\B{r}}, \quad \widehat{\B{r}}=\Theta_{\theta}^{T}\cdot \left[\theta' \B{r}\right].
\end{eqnarray}
Now, we are ready to show that the optimization problem \myref{weight-l1} can be solved explicitly by the shrinkage operator. To simplify the notation,
we denote $w=\B{r}_j^k+\B{q}^k/\mu$,
\begin{eqnarray}
&&\argmin_{\B{p}_j} \|\B{p}_j\|_1+\frac{\mu}{2}\|\B{r}_j^k+\B{q}^k/\mu-\Theta_{\theta_j}\B{p}_j\|_{2,\theta_j}^2\nonumber\\
&=&\argmin_{\B{p}_j} \|\B{p}_j\|_1+\frac{\mu}{2}\|\Theta_{\theta_j}\B{p}_j-P_{V(\theta_j)} \left(w\right)\|_{2,\theta_j}^2\nonumber\\
&=&\argmin_{\B{p}_j} \|\B{p}_j\|_1+\frac{\mu}{2}\|\Theta_{\theta_j}\cdot\left[\B{p}_j-\Theta_{\theta_j}^{T}\cdot \left[\theta_j'w\right]\right]
\|_{2,\theta_j}^2\nonumber\\
&=&\argmin_{\B{p}_j} \|\B{p}_j\|_1+\frac{\mu}{2}\|\B{p}_j-\Theta_{\theta_j}^{T}\cdot \left[\theta_j'w\right]\|_{2}^2\nonumber\\
&=&\mathcal{S}_{\mu^{-1}}\left(\Theta_{\theta_j}^{T}\cdot \left[\theta_j'\left(\B{r}_j^k+\B{q}^k/\mu\right)\right]\right),
\label{der-wavelet}
\end{eqnarray}
where $\mathcal{S}_{\tau}$ is the shrinkage operator defined below:
\begin{eqnarray}
  \mathcal{S}_{\tau}(x)=\mbox{sgn}(x)\max(|x|-\tau,0).
\end{eqnarray}

Notice that the matrix vector product in \myref{der-wavelet} has the following structure by the definition of $\Theta_{\theta_j}$ in \myref{def-Theta}
\begin{eqnarray*}
  \Theta_{\theta_j}^{T}\cdot \left(\theta_j'\B{r}\right)=\left[\Pi_{\theta_j}^T\cdot \left(\cos\theta_j\;\B{r}\;\theta_j'\right),
\Pi_{\theta_j}^T\cdot \left(\sin\theta_j\;\B{r}\;\theta_j'\right)\right]^T.
\end{eqnarray*}
This is nothing but the wavelet transform of $\sin\theta_j\;\B{r}$ and $\cos\theta_j\;\B{r}$ in the $\theta_j$-coordinate, since
the columns of $\Pi_{\theta_j}$ are standard wavelet basis in the $\theta_j$-coordinate.
Then this product can be computed efficiently by interpolating $\B{r}\cos\theta_j$
and $\B{r}\sin\theta_j$ to the uniform grid in the $\theta_j$ coordinate and employing the fast wavelet transform.

Summarizing the above discussion, we obtain Algorithm \ref{alg:sweep-ALM-wavelet} based on the fast wavelet transform to solve the optimization problem \myref{opt-mul-linear}:
\begin{algorithm}
\floatname{algorithm}{Algorithm}
\caption{(Sweeping ALM accelerated by the fast wavelet transform)}
\label{alg:sweep-ALM-wavelet}
\begin{algorithmic}[1]
\REQUIRE $a^0_{\theta_j}=b^0_{\theta_j}=0,\; j=1,\cdots, M,\; \B{q}^0=0$.
\WHILE{not converge}
\FOR{j=1:M}
\STATE Compute
\begin{eqnarray*}
\B{r}_j^n&=&f-\sum_{l=1}^{j-1}\left(a_{\theta_l}^{n+1}\cos\theta_l+b_{\theta_l}^{n+1}\sin\theta_l\right)
-\sum_{l=j+1}^{M}\left(a_{\theta_l}^{n}\cos\theta_l+b_{\theta_l}^{n}\sin\theta_l\right).
\end{eqnarray*}
\STATE Interpolate $\B{R}=\B{r}_j^n+\B{q}^n/\mu$ from $\{t_i\}_{i=1}^N$ in the
physical space  to a uniform mesh in the
 $\theta_j$-coordinate to get $\B{R}_{\theta_j}$ and compute the wavelet representations:
  \begin{eqnarray}
    R_{\theta_j,\,k}=\mbox{Interpolate}\;\left(\theta_j(t_i),\B{R},\theta_{j,\,k}\right),\nonumber
  \end{eqnarray}
where $\theta_{j,\,k}, \; j=0,\cdots,N-1$ are uniformly distributed in the
 $\theta_j$-coordinate, i.e. $\theta_{j,\,k}=2\pi L_{\theta_j}\; k/N$ and the interpolation is done by a cubic spline.
And 
compute
  \begin{eqnarray}
\label{eq:coe-wavelet-1}
    \widetilde{\B{a}}=\sum_{k=1}^N\Pi_{\theta_j}^T(\theta_{j,k})\cdot (\cos\theta_{j,k})R_{\theta_j,\,k},\quad\quad
    \widetilde{\B{b}}=\sum_{k=1}^N\Pi_{\theta_j}^T(\theta_{j,k})\cdot (\sin\theta_{j,k})R_{\theta_j,\,k}.
  \end{eqnarray}
This computation can be accelerated by the fast wavelet transform.

\STATE Apply the shrinkage operator compute $a_{\theta_j}$ and $b_{\theta_j}$ in the $\theta_j$-coordinate:
\begin{eqnarray}
\label{eq:coe-reconstruct-1}
a_{\theta_j}^{n+1}(\theta_{j,k})=\Pi_{\theta_j}(\theta_{j,k})\cdot \mathcal{S}_{\mu^{-1}}(\widetilde{\B{a}}),\quad\quad
b_{\theta_j}^{n+1}(\theta_{j,k})=\Pi_{\theta_j}(\theta_{j,k})\cdot \mathcal{S}_{\mu^{-1}}(\widetilde{\B{b}}).
\end{eqnarray}
This step can also be accelerated by the wavelet reconstruction algorithm.

\STATE Interpolate $a_{\theta_j}$ and $b_{\theta_j}$ back to
the physical grid points $\{t_i\}_{i=1}^N$ by a cubic spline.
\ENDFOR
\STATE Compute
\begin{eqnarray*}
\hspace{-4mm}\B{q}^{n+1}=\B{q}^n+\mu \left(f-\sum_{j=1}^M \left(a_{\theta_j}^{n+1}\cos\theta_j+b_{\theta_j}^{n+1}\sin\theta_j\right)\right).
\end{eqnarray*}
\ENDWHILE
\end{algorithmic}
\end{algorithm}

\begin{remark}
The continuous version of formula \myref{eq:coe-wavelet-1} is given as follows:
    \begin{eqnarray}
\widetilde{\B{a}}&=&\left\{ \left(\int R_{\theta_j}(t)\cos t\;\psi_{l,n}(t)\mathd t\right)_{l, n\in \mathbb{Z},\atop 0 < l\le l_0}
, \left(\int R_{\theta_j}(t)\cos t\;\varphi_{l_0,n}(t)\mathd t\right)_{n\in \mathbb{Z}}\right\},\\
\widetilde{\B{b}}&=&\left\{ \left(\int R_{\theta_j}(t)\sin t\;\psi_{l,n}(t)\mathd t\right)_{l, n\in \mathbb{Z},\atop 0 < l\le l_0}
, \left(\int R_{\theta_j}(t)\sin t\;\varphi_{l_0,n}(t)\mathd t\right)_{n\in \mathbb{Z}}\right\}
  \end{eqnarray}
where $R_{\theta_j}(t)=R(\theta_j^{-1}(t))$.

Obviously,  $\widetilde{\B{a}}, \widetilde{\B{b}}$ are just coefficients of
$R_{\theta_j}(t)\cos t, R_{\theta_j}(t)\sin t$ over $\left(\psi_{l,n}\right)_{l, n\in \mathbb{Z},\atop 0 < l\le l_0},
\left(\varphi_{l_0,n}\right)_{n\in \mathbb{Z}}$.

Similarly, the computation in step 5 can be accelerated by a fast wavelet reconstruction algorithm, since each
column of the matrix $\Pi_{\theta_j}$ is standard wavelet basis in the $\theta_j$-space.
\end{remark}


\section{Generalization for signals with outliers}
One advantage of the formulation \myref{opt-l1} is that it can be generalized to deal with more complicated data with some minor modifications. In this
section, we will give one generalization for signals with outliers.

In order to deal with this kind of signals, we have to enlarge the dictionary since the outliers are not sparse over the time-frequency dictionary.
Fortunately, we know that the outliers are sparse over the basis consisting of the impulses $\delta[n-i], \; i=1,\cdots,N$, where $N$ is the number of
samples and
\begin{eqnarray}
  \delta[n]=\left\{\begin{array}{cc}1,& n=0,\\
0, & n\neq 0.
\end{array}\right.
\end{eqnarray}
If we enlarge the dictionary to include all the impulses, then the generalized formulation can be used to decompose signals with outliers.

More specifically, in this case, the optimization problem is formulated in the
following way,
 \begin{eqnarray}
  \label{opt-outlier}
&&  \min_{\B{x}, \theta_1\cdots, \theta_M}  \|\B{x}\|_1+\|\B{z}\|_1,\\
&& \mbox{subject to:}\quad \Phi_{\theta_1,\cdots,\theta_M}\cdot \B{x}+\B{z}=f,\nonumber
\end{eqnarray}
where $\Phi_{\theta_1,\cdots,\theta_M}$ is given in \myref{fourier-theta}.

Using the idea similar to that of Algorithm 1, we obtain the following Gauss-Newton type method for the above optimization problem \myref{opt-outlier}:
\begin{algorithm}[H]
\floatname{algorithm}{Algorithm}
\caption{(Gauss-Newton type iteration with outliers)}
\label{alg:gauss-newton-outlier}
\begin{algorithmic}[1]
\REQUIRE Initial guess of phase functions $\theta_j^0,\; j=1,\cdots,M$, $\eta=l_0$, where $l_0$ is same as that in \myref{wavelet-envelope-j}.
\ENSURE Phase functions and the corresponding envelopes: $\theta_j,\;a_j,\quad j=1,\cdots,M$.
\WHILE{$\eta\ge 1$}
\WHILE{ $\D \sum_{j=1}^M\|\theta_j^{n+1}-\theta_j^{n}\|_2>\epsilon_0$}
\STATE Solve the following $l^1$ optimization problem:
\begin{eqnarray*}
&&\hspace{-1cm}\left(\B{\widetilde{a}}^{n+1},\B{\widetilde{b}}^{n+1},\B{z}^{n+1}\right)=
\argmin_{\B{x},\B{y},\B{z}}  (\|\B{x}\|_1+\|\B{y}\|_1+\|\B{z}\|_1), \\
&&\hspace{-1cm}\mbox{subject to:}\quad \Phi_{\theta_1^n,\cdots,\theta_M^n}\cdot\B{x}
+\Psi_{\theta_1^n,\cdots,\theta_M^n}\cdot\B{y}+\B{z}=f.\nonumber
\end{eqnarray*}
\STATE Calculate the envelopes in the same way as we did in Algorithm \ref{alg:guass-newton}.
\STATE  Update $\theta^n_j,\; j=1,\cdots,M$ in the same way as we did in Algorithm \ref{alg:guass-newton}.
\ENDWHILE
\STATE $\eta=\eta-1$.
\ENDWHILE
\end{algorithmic}
\end{algorithm}
Moreover, the $l^1$ optimization problem in the above iterative algorithm can be solved by the following sweeping ALM method:
\begin{algorithm}[H]
\floatname{algorithm}{Algorithm}
\caption{(Sweeping ALM with outliers)}
\begin{algorithmic}[1]
\REQUIRE $\B{p}_j^0=0,\;j=1,\cdots,M$, $\B{q}^0=0$, $\mu>0$.
\ENSURE Phase functions and the corresponding envelopes: $\theta_j,\;a_j,\quad j=1,\cdots,M$.
\WHILE{not converge}
\FOR{ $j=1:M$}
\STATE $\D\B{r}_j^k=f-\sum_{l=1}^{j-1}\Theta_{\theta_l}\B{p}_l^{k+1}-\sum_{l=j+1}^{M}\Theta_{\theta_l}\B{p}_l^{k}-\B{z}^k$.
\STATE $\D \B{p}_j^{k+1}=\arg\min_{\B{p}_j} \|\B{p}_j\|_1+\frac{\mu}{2}\|\B{r}_j^k+\B{q}^k/\mu-\Theta_{\theta_j}\B{p}_j\|_2^2$.
\ENDFOR
\STATE $\D\B{r}_0^k=f-\sum_{l=1}^{M}\Theta_{\theta_l}\B{p}_l^{k+1}$.
\STATE $\D\B{z}^{k+1}=\mathcal{S}_{\mu^{-1}}(\B{r}_0^k+\B{q}^k/\mu)$.
\STATE $\D\B{q}^{k+1}=\B{q}^k+\mu \left(f-\sum_{j=1}^M \Theta_{\theta_j}\B{p}_j^{k+1}-\B{z}^{k+1}\right)$.
\ENDWHILE
\end{algorithmic}
\end{algorithm}
The computation of $\D \B{p}_j^{k+1}=\argmin_{\B{p}_j} \|\B{p}_j\|_1+\frac{\mu}{2}\|\B{r}_j^k+\B{q}^k/\mu-\Theta_{\theta_j}\B{p}_j\|_2^2$ can be accelerated by Algorithm \ref{alg:sweep-ALM-wavelet} in the previous section.

\section{Numerical Results}
\label{sec:numerics}

In this section, we present several numerical results to
demonstrate the effectiveness of our time-frequency
analysis methods. In our previous paper \cite{HS12}, we have performed extensive numerical experiments to demonstrate the effectiveness of our data-driven time-frequency analysis method for signals with good scale separations. To save space, we will not consider the examples with good scale separation in this paper and consider more chanllenging signals that do not have good scale separation property.

\begin{figure}

    \begin{center}

\includegraphics[width=0.28\textwidth]{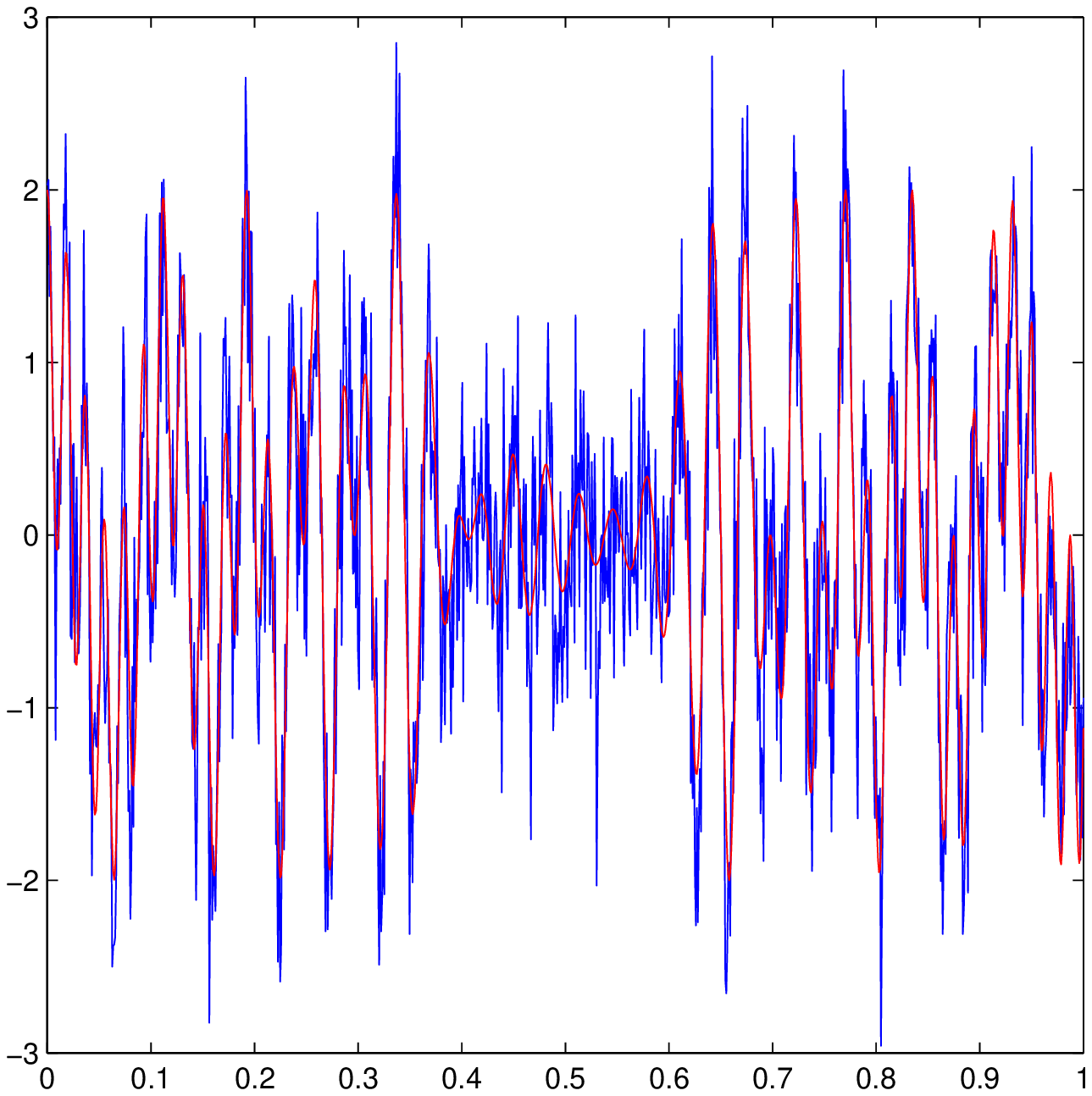}
\includegraphics[width=0.28\textwidth]{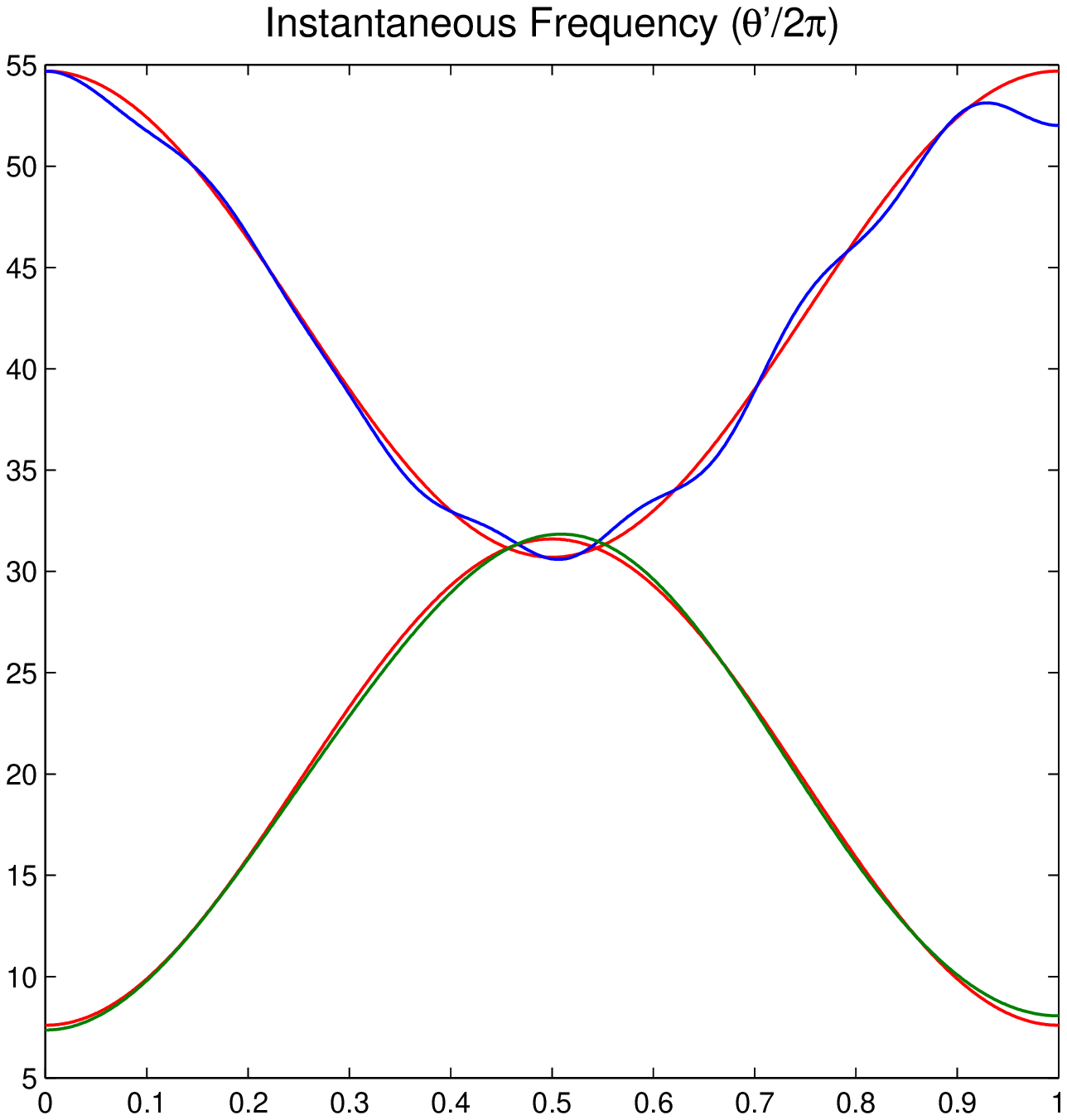}
\includegraphics[width=0.34\textwidth]{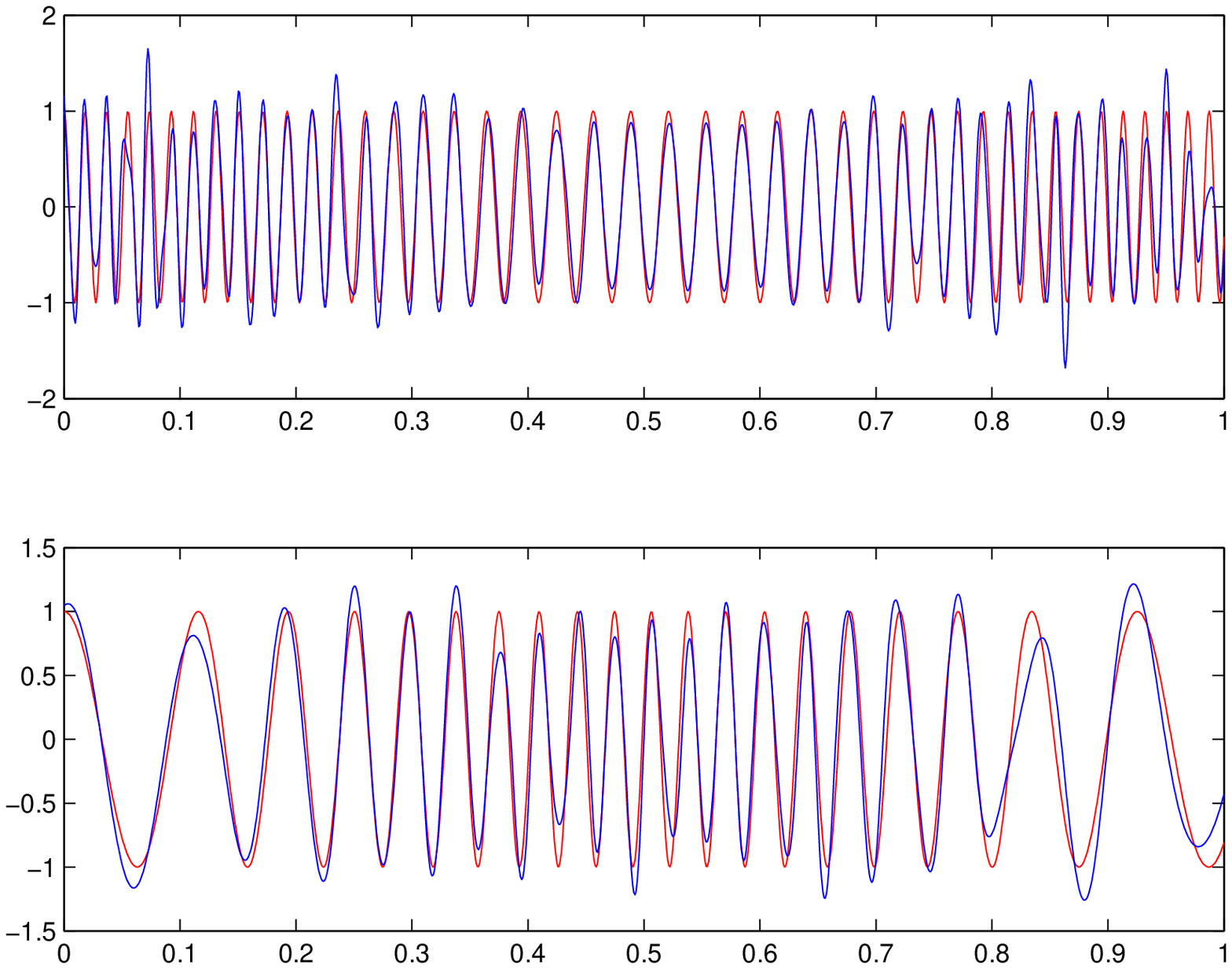}
     \end{center}
    \caption{  \label{Fig-result-ex2} Left: Original signal in Example 2, red curve is the clean signal without noise;
Middle: instantaneous frequencies, red: exact, blue: numerical; Right: corresponding IMFs, red: exact, blue: numerical.}
\end{figure}

\vspace{3mm}
\noindent \textbf{Example 1:}
In the first example, the signal contains two different components
whose instantaneous frequencies intersect with each other. More specifically,
the signal is generated by the formula below:
\begin{eqnarray}
\label{data-ex1}
  f=\cos\theta_1(t)+\cos\theta_2(t)+X(t),\;t\in [0,1],
\end{eqnarray}
where the phase function $\theta_1,\theta_2$ are given as following:
\begin{eqnarray}
  \label{eq:theta1}
\theta_1(t)&=&39.2\pi t-12\sin 2\pi t,\\
\label{eq:theta2}
\theta_2(t)&=&85.4\pi t+12\sin 2\pi t.
\end{eqnarray}
and $X(t)$ is white noise with zero
mean and variance $\sigma^2=1$.
The signal is sampled over 1024 grid points which are uniformly distributed over the
interval $[0,1]$. The original signal is shown in Fig. \ref{Fig-result-ex2}.

For this signal,
the classical time-frequency analysis methods, such as
the windowed Fourier transform, the wavelet transform give a poor result
near the intersection where the two instantaneous frequencies cross each other. The EMD method and the data-driven time-frequency analysis method
introduced in our previous paper \cite{HS12} also have problem near the intersection.

The result given by the data-driven time-frequency analysis method proposed in this paper is shown in Fig. \ref{Fig-result-ex2}.
In the computation, the initial guesses of the instantaneous frequencies
are chosen to be $128\pi t$ and $32\pi t$ respectively, which are far from the ground truth. As we can see, even with these rough initial guesses,
our iterative algorithm still can recover the instantaneous frequencies and the corresponding IMFs with reasonable accuracy,
although the signal is polluted by noise. We note that the end-effect is more pronounced in this case due to the noise pollution.




\begin{figure}
    \begin{center}
\includegraphics[width=0.28\textwidth]{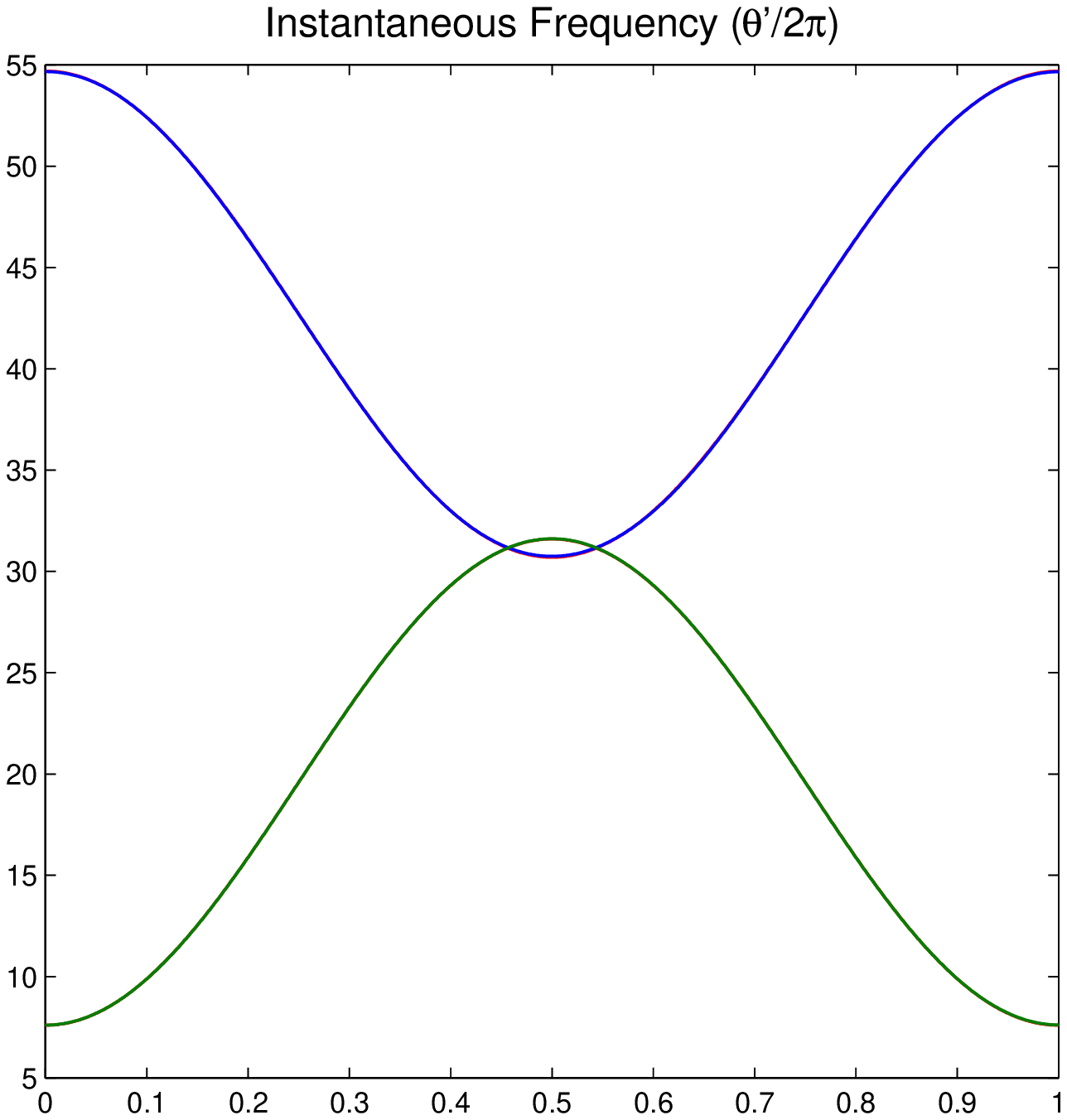}
\includegraphics[width=0.34\textwidth]{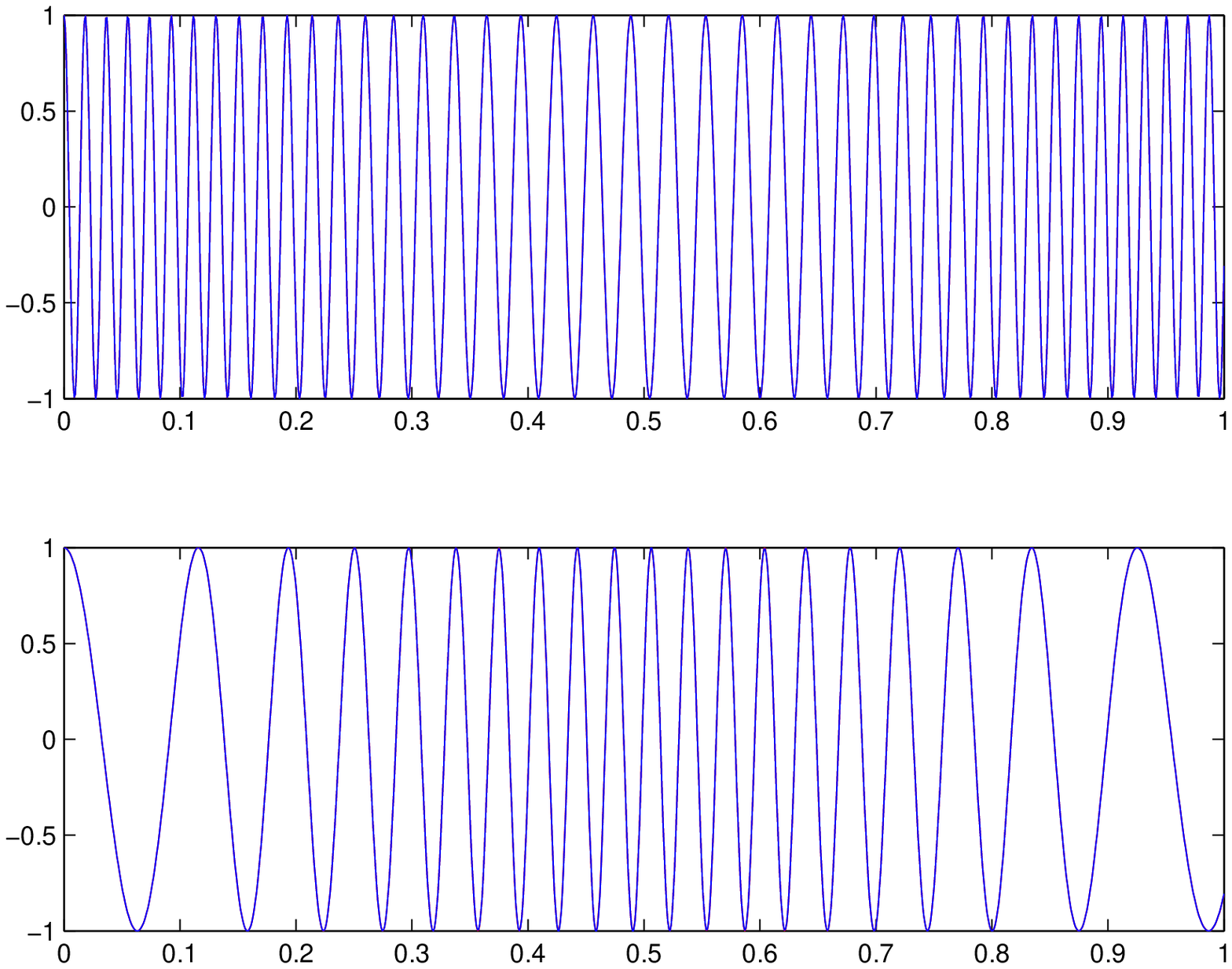}
\includegraphics[width=0.28\textwidth]{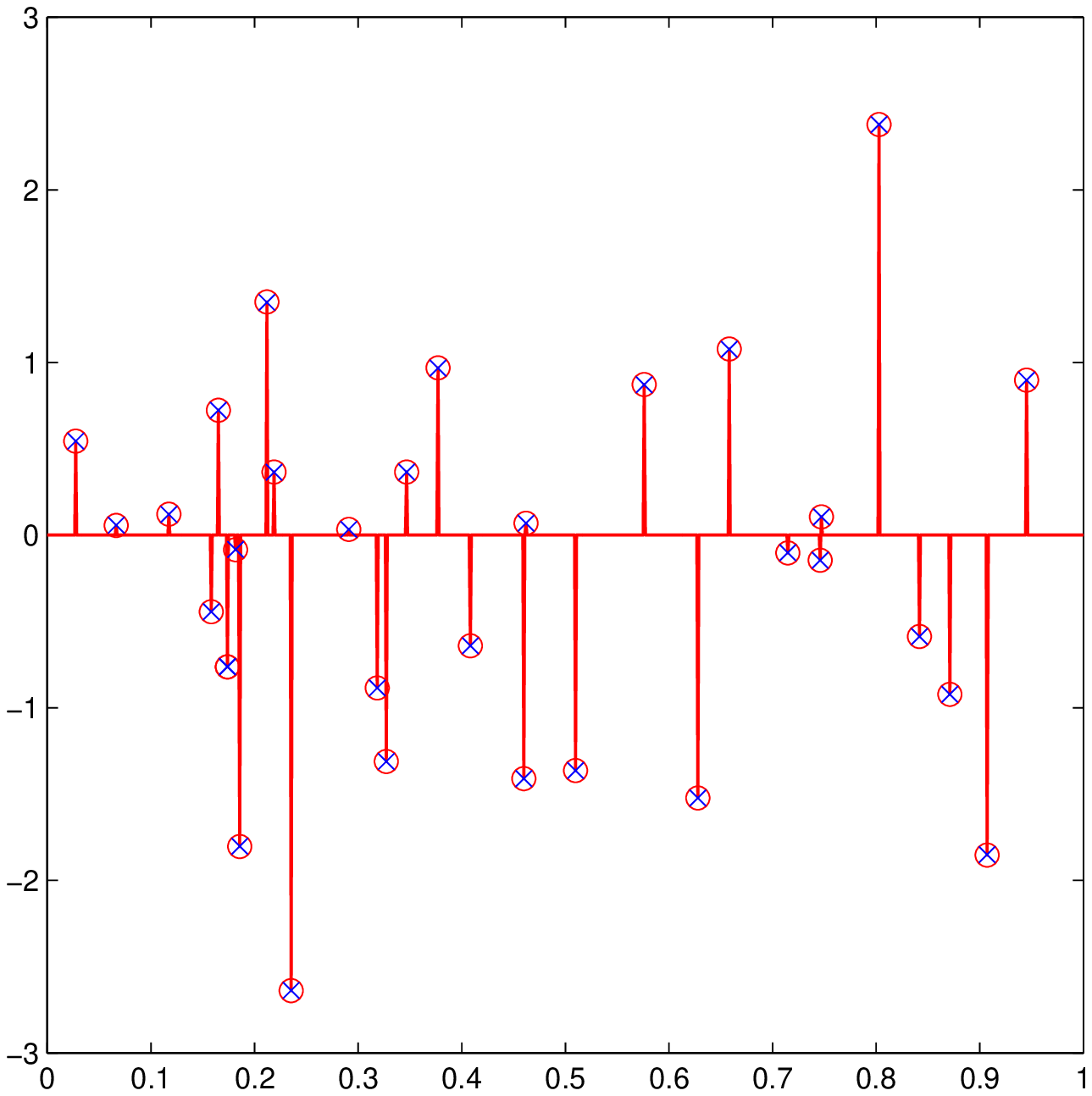}
\end{center}
     \caption{  \label{Fig-result-ex4} Left: instantaneous frequencies, red: exact, blue: numerical;
Middle: corresponding IMFs, red: exact, blue: numerical; Right: Outliers, red circle: exact, blue cross: numerical.}
\end{figure}

\vspace{3mm}
\noindent
\textbf{Example 2:}
The next signal that we consider is polluted by outliers. We generate the signal by using the following formula:
\begin{eqnarray}
\label{data-ex4}
  f=\cos\theta_1(t)+\cos\theta_2(t)+\sigma(t).
\end{eqnarray}
The signal is sampled over 1024 uniform grid points. Among these samples, there are 32 samples that are outliers.
The locations of these outliers are selected randomly and the strengths satisfy the normal distribution.
In Fig. \ref{Fig-result-ex4}, we present the results that we obtain using
the algorithm given in Section 5. As we can see, both the IMFs and the outliers are captured very accurately.

\begin{figure}
    \begin{center}
\includegraphics[width=0.28\textwidth]{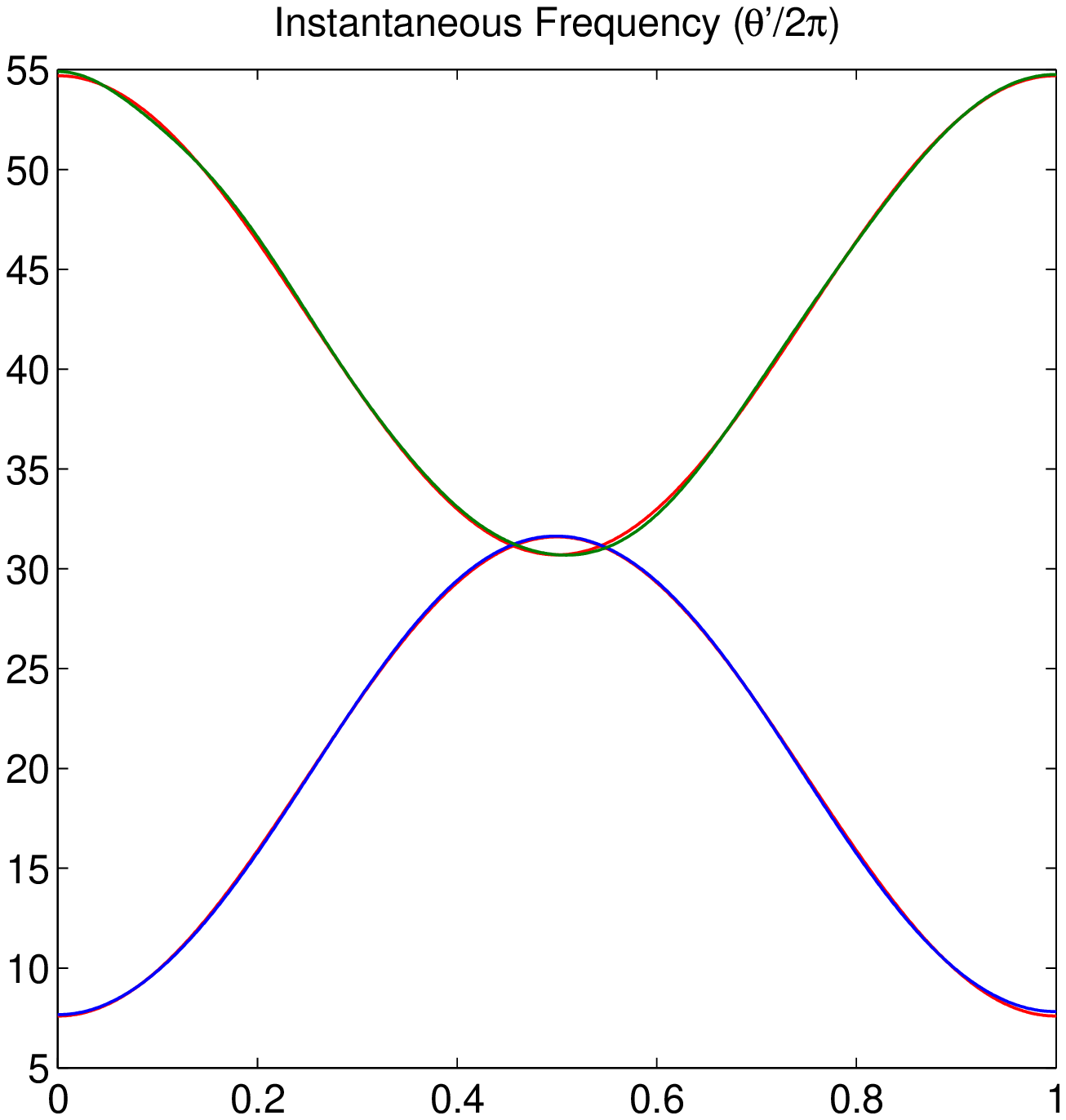}
\includegraphics[width=0.34\textwidth]{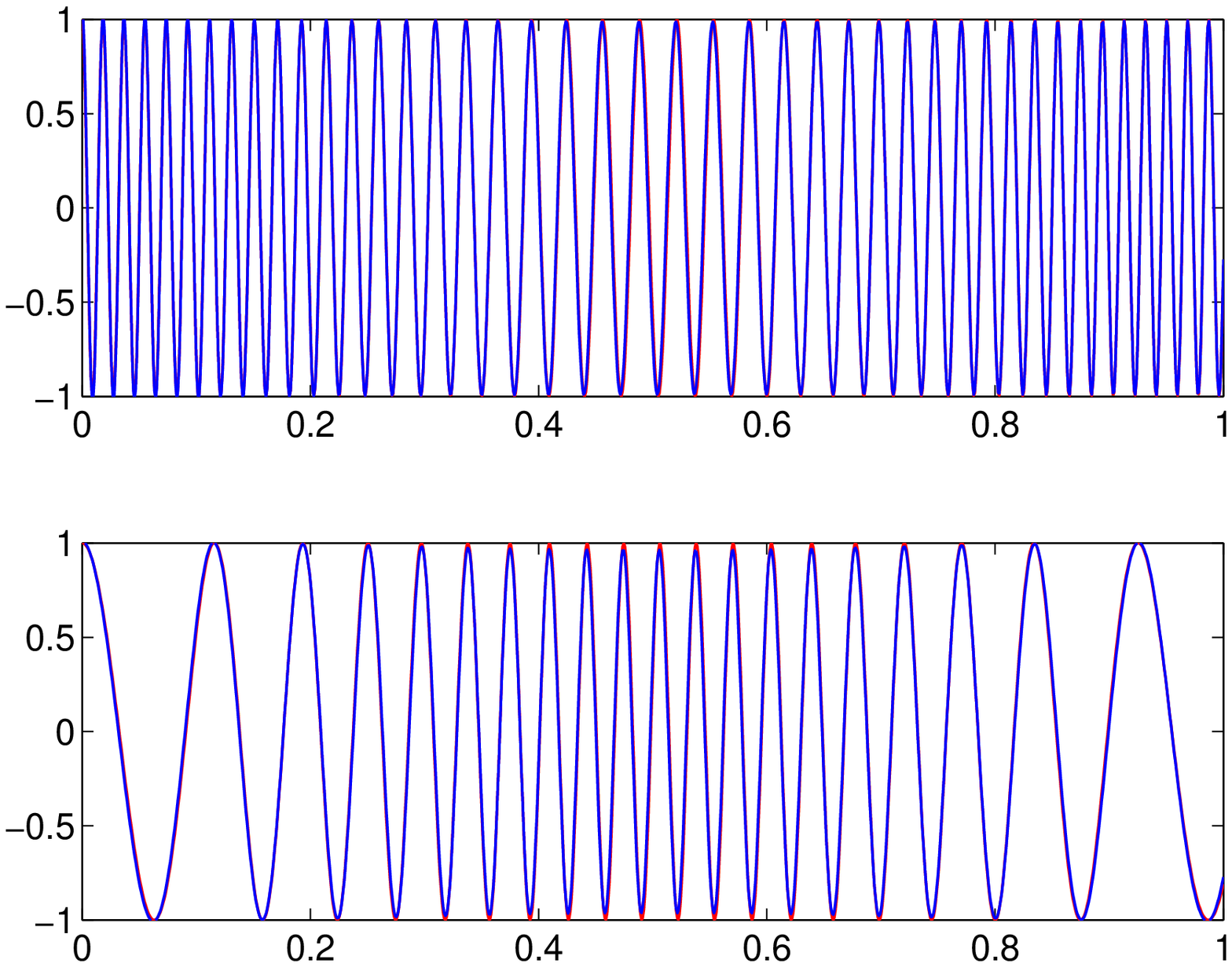}
\includegraphics[width=0.28\textwidth]{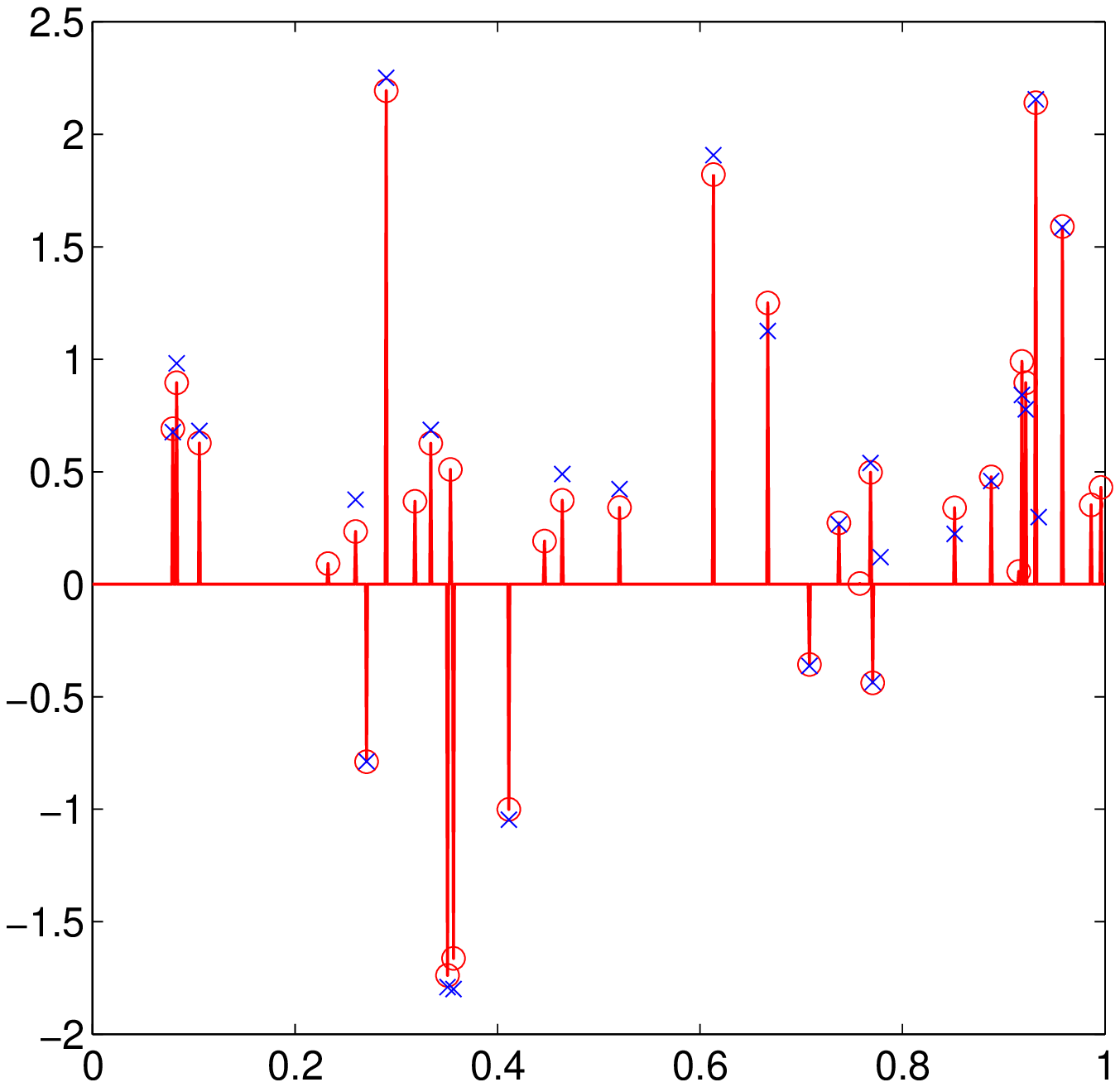}
\end{center}
     \caption{  \label{Fig-result-ex4-noise} Left: instantaneous frequencies, red: exact, blue: numerical;
Middle: corresponding IMFs, red: exact, blue: numerical; Right: Outliers, red circle: exact, blue cross: numerical.}
\end{figure}

We also test the signal with outliers and noise. The results is shown in Fig. \ref{Fig-result-ex4-noise}. In this example,
the noise and the outliers are added to the original signal together.
The signal is sampled over 1024 uniform grid points. Among these samples, there are 32 samples that are outliers.
The locations of these outliers are selected randomly and the strengths satisfy the normal distribution whose standard deviation is 1.
The noise is Gaussian noise and the standard deviation is 0.1.

In this case, the instantaneous frequencies and the IMFs are still quite accurate. But the outliers are not captured as well as in the case with no noise. We would like to emphasize that it would be hard to distinguish the outliers from the noise when the amplitude of the outliers is small.
For the outliers whose amplitude is large, they can be separated from the noise by a proper shrinkage operator. However, the shrinkage operator also kills the outliers whose whose amplitude is comparable to or smaller than the noise level.

\vspace{3mm}
\noindent \textbf{Example 3:}

In our third example, we consider a real signal, a bat chirp signal. It is the digitized
 echolocation pulse emitted by the Large Brown Bat, Eptesicus Fuscus
\footnote{The authors wish to thank Curtis Condon, Ken White, and Al Feng
of the Beckman Institute of the University of Illinois for the bat data and for permission to use it in this paper.}.
The signal includes 400 samples and the sampling period is 7 microseconds, so the total time span is 2.8 miliseconds.

\begin{figure}
    \begin{center}
\includegraphics[width=0.28\textwidth]{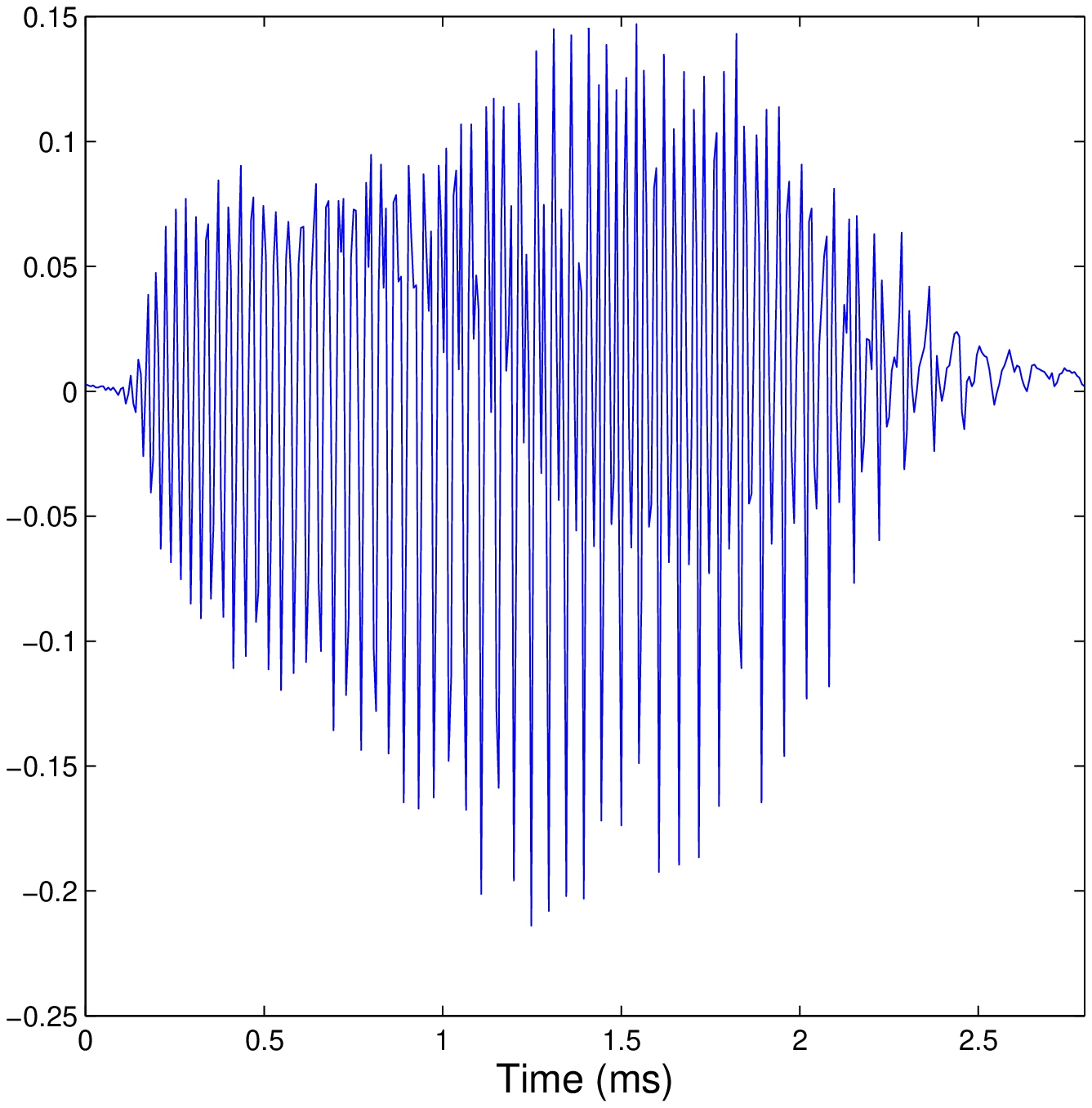}
\includegraphics[width=0.35\textwidth]{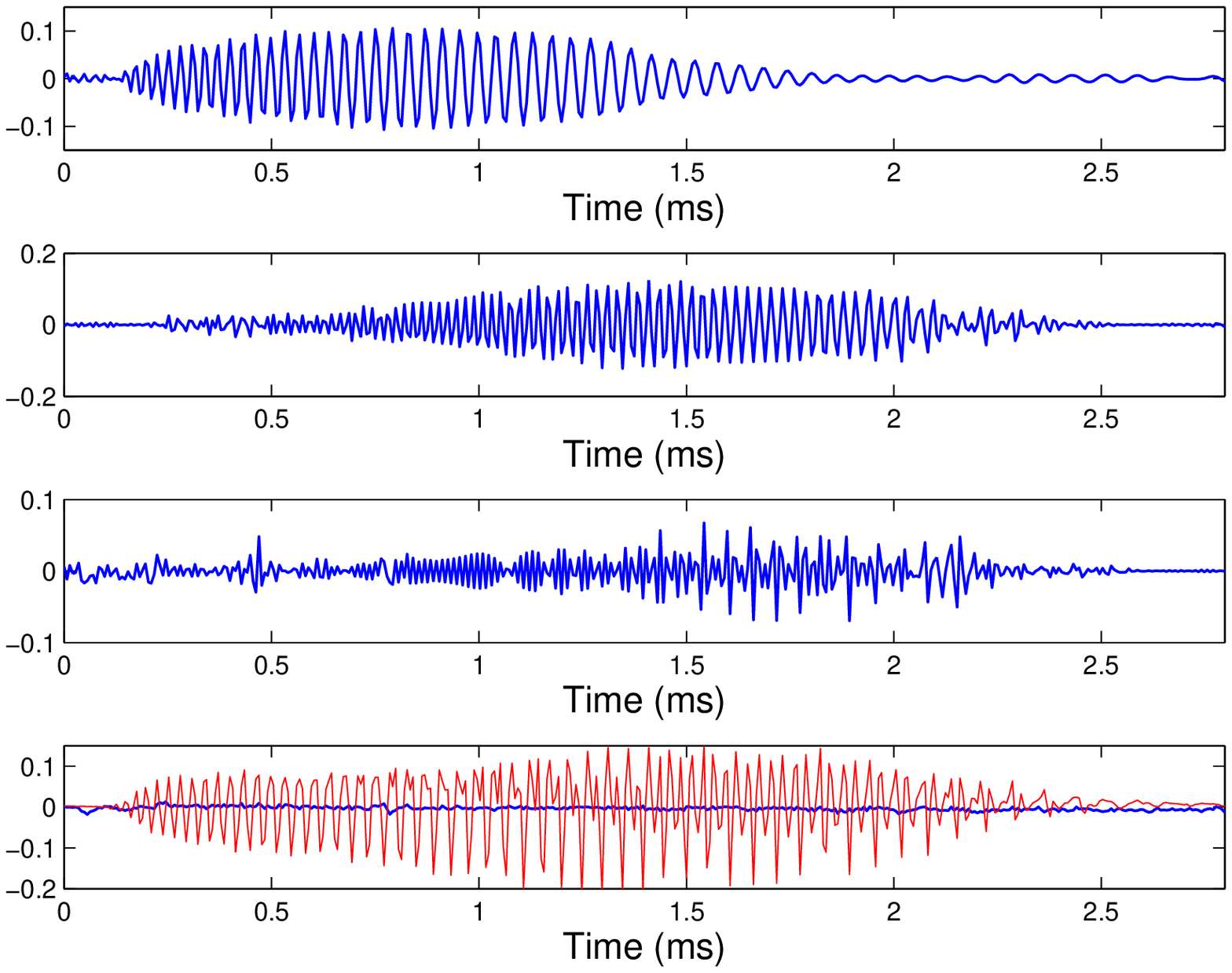}
\includegraphics[width=0.29\textwidth]{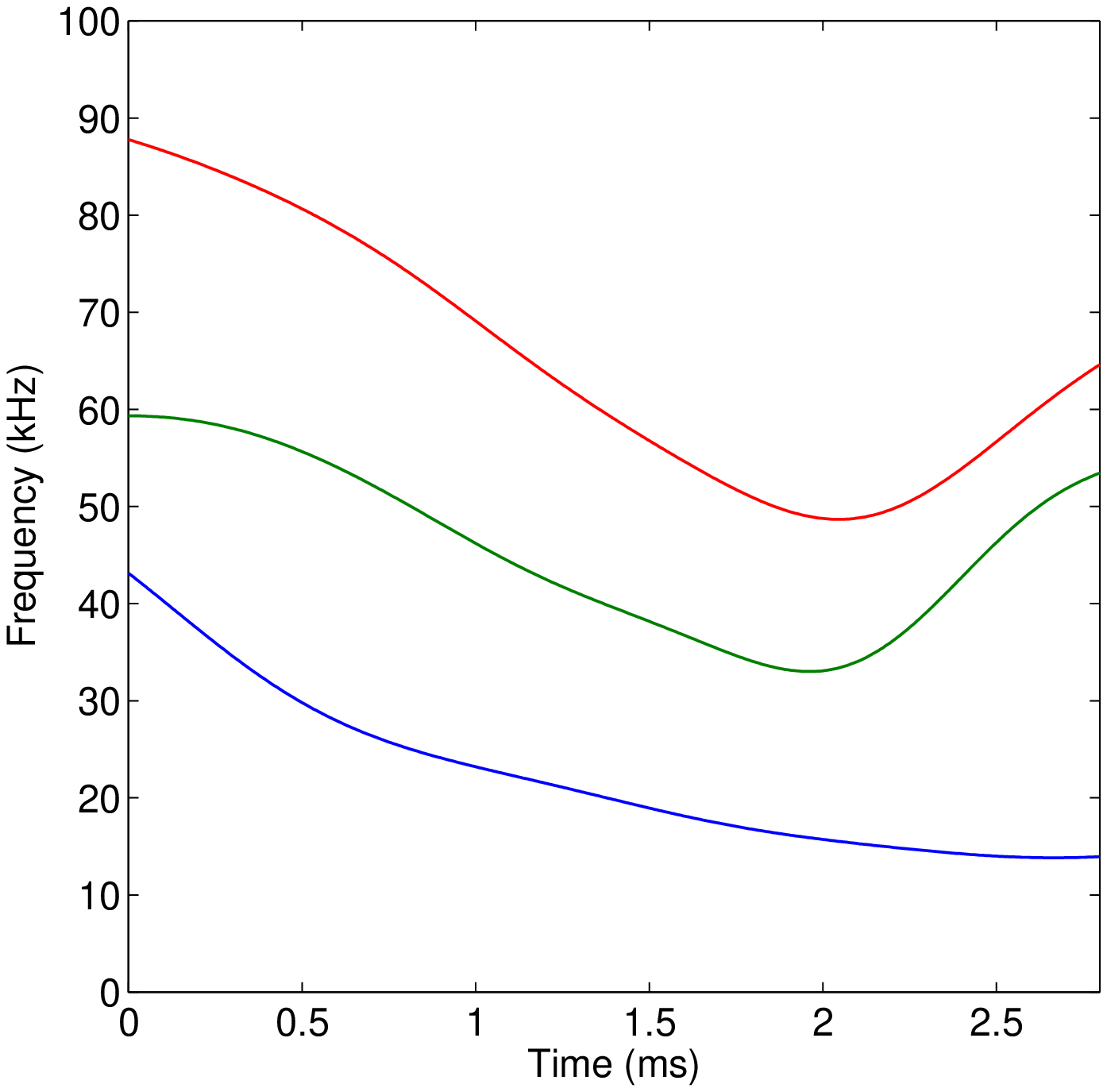}
     \end{center}
    \caption{  \label{Fig-bat}Left: Bat chirp signal;
Middle: IMFs; Right: Instantaneous frequencies.}
\end{figure}

The signal is shown in the left panel of Fig. \ref{Fig-bat}. The IMFs and instantaneous frequencies obtained by our method
are given in the middle and right panel of Fig. \ref{Fig-bat}. Our method could gives precise instantaneous frequencies and
also the sparse decomposition of the original signal. From Fig. \ref{Fig-bat}, we can see that the signal is approximated very
well by only three IMFs. Near the boundaries, the original signal and the IMFs are all very small. In these regions,
the frequencies actually do not have any physical meaning. They are only auxiliary variables in the algorithm.
In the region in which the amplitude of the signal is order one, the recovered instantaneous frequencies reveal some interesting patterns that have not been seen before using traditional time-frequency methods. The physical significance of these patterns need to be further investigated in the future.

\vspace{3mm}
\noindent \textbf{Example 4:}

In the last example, we consider the data from an ODE system. We consider a multiple degree of freedom (MDOF) system.
\begin{eqnarray}
\label{eqn:MDOF}
  \ddot{\mathbf{u}}+K(t) \mathbf{u}=0
\end{eqnarray}
where $K$ is an $N\times N$ symmetric positive definite stiffness matrix and $\mathbf{u}$ is an $N\times 1$ vector, which typically represents the displacement of certain engineering structure. This kind of ODE system is widely used to model the movements of structures \cite{Chopra}, such as buildings, bridges etc. In many applications, we
want to recover the stiffness matrix $K$ (at least part of it) from incomplete measurement of the solution $\mathbf{u}$.

Here, we assume that $K(t)$ is slowly varying. Under this assumption, the solutions, $u_j(t), j=1,\cdots, N$, have the following approximate expression:
\begin{eqnarray}
\label{eqn:sol-MDOF}
  u_j(t)=\sum_{k=1}^N \rho_{j,k} e^{i\theta_k(t)} ,
\end{eqnarray}
and $(\theta_k'(t))^2, k=1,\cdots,N$ are eigenvalues of $K(t)$. Using this formulation, we can see that the instantaneous frequencies could help us to
retrieve the stiffness matrix.

As a test example, we consider a simple case where the degree of freedom $N=2$.
\begin{eqnarray}
  K=\left[\begin{array}{cc}
k_1+k_2& -k_2\\
-k_2&k_2+k_3
\end{array}
\right]
\end{eqnarray}
and
\begin{eqnarray}
  k_1=k_3=100\cos(0.2\pi t)+500,\quad k_2=400\cos(0.2\pi t)+400
\end{eqnarray}
The initial conditions are
\begin{eqnarray}
  u_1(0)=1,\quad u_1'(0)=0,\quad u_2(0)=2,\quad u_2'(0)=0.
\end{eqnarray}
This system models the movement of two objects of equal masses connected by springs. And $k_1, k_2, k_3$ are stiffness values of springs.

\begin{figure}
    \begin{center}
\includegraphics[width=0.45\textwidth]{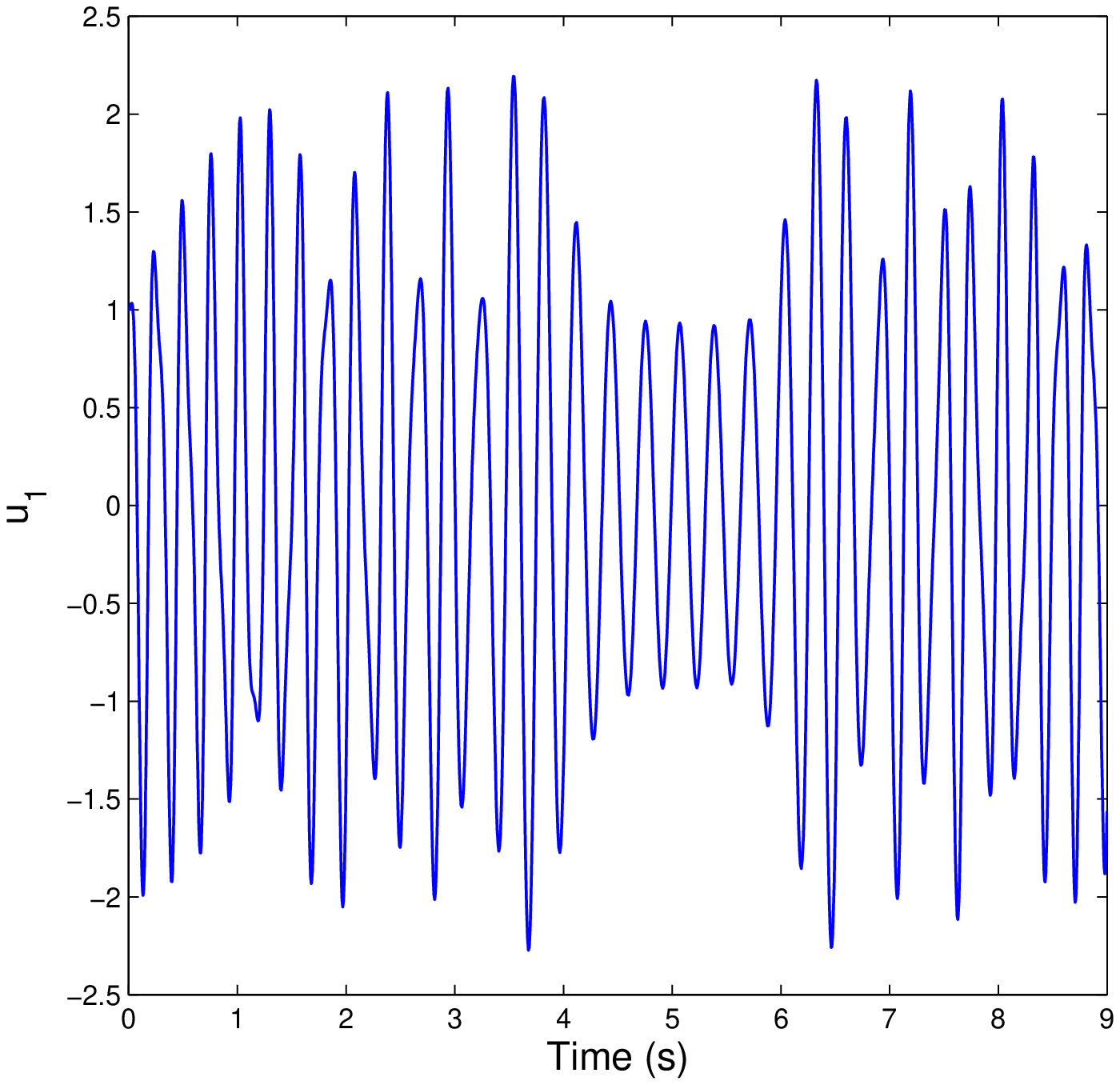}
\includegraphics[width=0.43\textwidth]{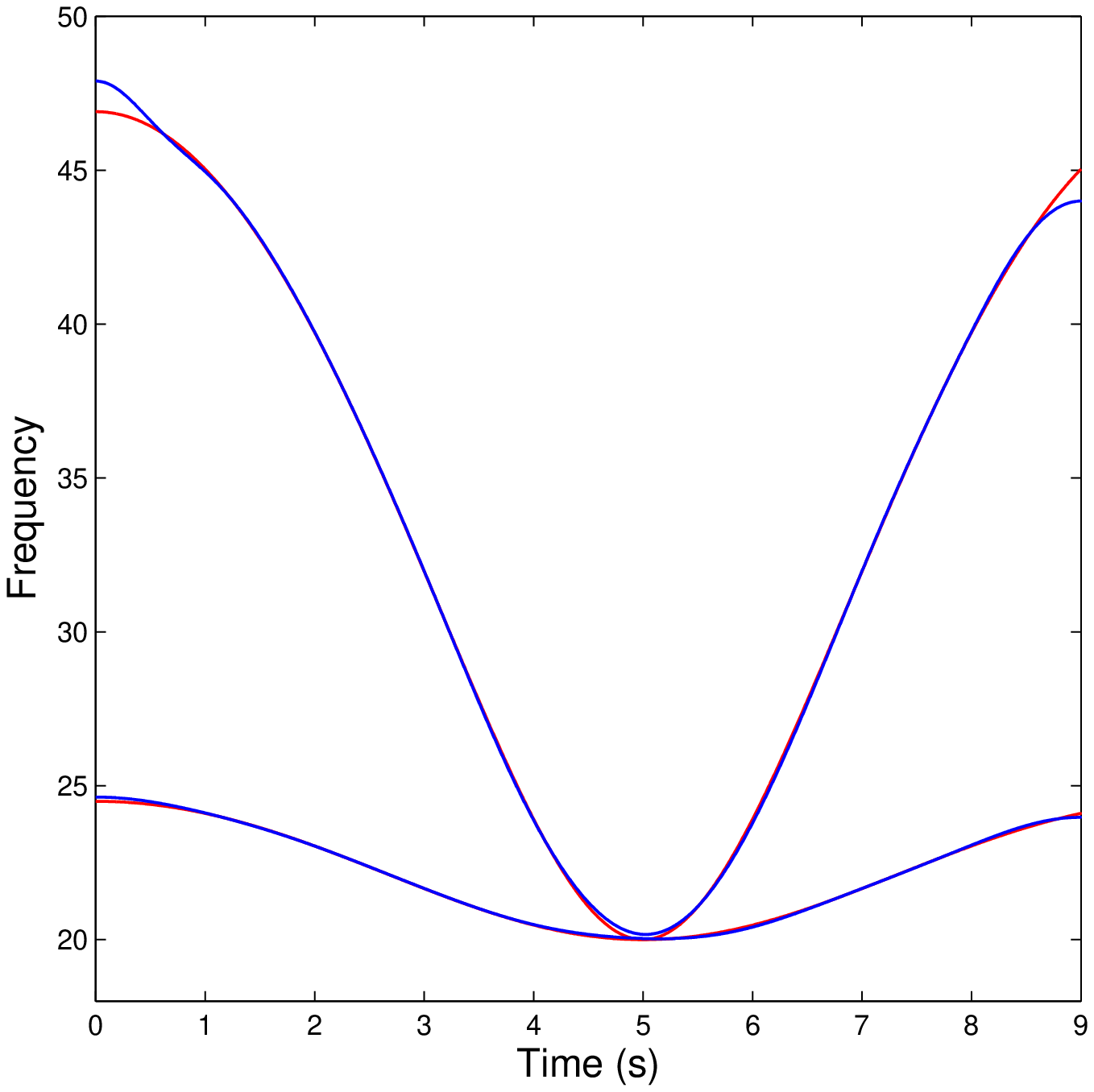}
     \end{center}
    \caption{  \label{Fig-MDOF}Left: Solution $u_1$ of the ODE system \eqref{eqn:MDOF} ;
Right: Instantaneous frequencies, red: theoretical frequencies; blue: numerical results.}
\end{figure}

The above system is solved from 0 to 9. The initial guesses of the phase functions are $20t$ and $40t$.
Here we only analyze the first component of the solution $u_1(t)$, which is shown in the left figure of Fig. \ref{Fig-MDOF}.
According to \eqref{eqn:sol-MDOF}, the theoretical frequencies are $\sqrt{(k_1+k_3)/2}$ and $\sqrt{(k_1+k_3+4k_2)/2}$. In the right figure of Fig. \ref{Fig-MDOF},
we compare the theoretical frequencies and the numerical results given by our method. As we can see, they match very well even near the intersection point.
This toy example shows that our method indeed has the capability to retrieve some information of the physical process hidden within the signal. 
Now we are trying to apply our method to
analyze the signals from real bridges.

\section{Concluding remarks}

In this paper, we introduced a noval formulation to obtain sparse Time-Frequency decomposition and the corresponding instantaneous frequencies.
We formulated the decomposition as a dictionary learning problem. The dictionary is parametrized by phase functions
and the phase functions are determined by the signal itself. Based on our previous work and the methods of dictionary learning,
we developed an iterative algorithm to look for the phase functions and the corresponding decomposition.
By designing the dictionary carefully to make them orthogonal in
the coordinate of the phase functions, we can accelerate the algorithm by
using the fast wavelet transform. This makes our algorithm very efficient.

Another advantage of this method is that it can be easily generalized to deal with more complicated data that are not sparse over the time-frequency dictionary,
such as data with outliers. For this kind of signals, we just need to enlarge the dictionary and follow the similar procedure
to look for the sparsest decomposition over this enlarged dictionary. We presented several numerical examples to demonstrate
the effectiveness of our method, including data that do not have scale separation and data that are polluted by noise or outliers.
The results that we obtained seem to suggest that our method can offer an effective way to decompose multiscale data
even with poor scale separation property.

We remark that decomposing several IMFs simultaneously increases the complexity of the optimization problem. As a result, the robustness
of this new method is not as good as the previous one that we introduced in \cite{HS12}. Moreover, to apply this method, we need to know some
information of the signal. We are considering
to combine with other time-frequency analysis method, such as synchrosqueezed wavelet transform \cite{DLW11} in our future work.

Another interesting problem that we are considering is to decompose data with intra-wave frequency modulation. This type of data is known to be very challenging. Naive application of traditional data analysis methods tends to introduce artificial harmonics. To deal with this kind of data, we have to extend the definition of the dictionary by replacing the cosine function by an unknwon periodic shape function that we need to find as part of the optimization problem. This work will be reported in our future paper. The other project we are working on is to apply the method in this paper to analyze the signals from real bridges and we have got some progress. More results will be reported in future.

\vspace{0.2in}
\noindent
{\bf Acknowledgments.}
This research was supported in part by NSF Grants No. DMS-318377 and DMS-1159138,  DOE Grant DE-FG02-06ER25727, and AFOSR MURI Grant FA9550-09-1-0613.  The research of Dr. Z. Shi was supported by a NSFC Grant 11201257.


\begin{thebibliography}{99}

\bibitem{AEB06}
M. Aharon, M. Elad, and A. M. Bruckstein. The K-SVD: An algorithm for designing
of overcomplete dictionaries for sparse representations.
{\it IEEE Transactions on
Signal Processing}
, \textbf{54(11)}, pp. 4311-4322, 2006.



\bibitem{Bert82}
D.P. Bertsekas, Constrained Optimization and Lagrange Multiplier Method, Academic Press, 1982.


\bibitem{BDL09} A. M. Bruckstein, D. L. Donoho, M. Elad, From sparse
solutions of systems of equations to sparse modeling of signals and
images, \textit{SIAM Review}, \textbf{51}, pp. 34-81, 2009.




\bibitem{Chopra} Anil K. Chopra, Dynamics OF Structures: Theory and Applications to
    Earthquake Engineering, Prentice-Hall, 1995.

\bibitem{Candes-Tao06} E. Cand$\grave{\mbox{e}}$s and T. Tao, Near
optimal signal recovery from random projections: Universal encoding
strategies?, \textit{IEEE Trans. on Information Theory}, \textbf{52(12)},
pp. 5406-5425, 2006.

\bibitem{CRT06a} E. Candes, J. Romberg, and T. Tao, Robust uncertainty
principles: Exact signal recovery from highly incomplete frequency
information, \textit{IEEE Trans. Inform. Theory}, \textbf{52}, pp.
489-509, 2006.

\bibitem{CRT06b} E. Candes, J. Romberg, and T. Tao, Stable signal
recovery from incomplete and inaccurate measurements, \textit{Comm.
Pure and Appl. Math.}, \textbf{59}, pp. 1207-1223, 2006.


\bibitem{CDS98} S. Chen, D. Donoho and M. Saunders, Atomic decomposition
by basis pursuit, \textit{SIAM J. Sci. Comput.}, \textbf{20}, pp.
33-61, 1998.

\bibitem{Daub92} I. Daubechies, Ten Lectures on Wavelets, CBMS-NSF
Regional Conference Series on Applied Mathematics, Vol. 61, SIAM Publications,
1992.

\bibitem{DLW11} I. Daubechies, J. Lu and H. Wu, Synchrosqueezed wavelet
transforms: an empirical mode decomposition-like tool, \textit{Appl.
Comp. Harmonic Anal.}, \textbf{30}, pp. 243-261, 2011.

\bibitem{Donoho06} D. L. Donoho, Compressed sensing, \textit{IEEE
Trans. Inform. Theory}, \textbf{52}, pp. 1289-1306, 2006.

\bibitem{DZ13} K. Dragomiretskiy and D. Zosso, Varitational Mode Decomposition, \textit{IEEE Trans. on Signal Processing}.

\bibitem{ESH07}
K. Engan, K. Skretting and
J. Hus{\o}y,
Family of iterative LS-based dictionary learning algorithms, ILS-DLA, for sparse signal representation,
{\it Digital Signal Processing}
, \textbf{17(1)}, pp. 32-49, 2007.

\bibitem{Gilles13} J. Gilles, Empirical Wavelet Transform, \textit{IEEE Trans. on Signal Processing}, \textbf{61}, pp. 3999-4010, 2013

\bibitem{GO09} Tom Goldstein and Stanley Osher, The Split Bregman method for $L_1$-regularized problems, {\it
SIAM J. Imaging Sci.}, {\bf 2}, pp. 323-343, 2009.


\bibitem{HS11} T. Y. Hou and Z. Shi, Adaptive data analysis via sparse
time-frequency representation, \textit{Advances in Adaptive Data Analysis},
\textbf{3}, pp. 1-28, 2011.

\bibitem{HS12} T. Y. Hou and Z. Shi, Data-driven time-frequency analysis,
\textit{Applied and Comput. Harmonic Analysis}, \textbf{35(2)}, pp. 284-308, 2013.

\bibitem{HST13}
T. Y. Hou, Z. Shi, P. Tavallali, Convergence of a data-driven time-frequency analysis method, submitted to
\textit{Applied and Comput. Harmonic Analysis}.

\bibitem{Huang98} N. E. Huang et al., The empirical mode decomposition
and the Hilbert spectrum for nonlinear and non-stationary time series
analysis, \textit{Proc. R. Soc. Lond. A}, \textbf{454} (1998), pp.
903-995.


\bibitem{LS00}
M. Lewicki and T. Sejnowski, Learning overcomplete representations.
{\it Neural Computation}, \textbf{12(2)}, pp. 337-365, 2000.

\bibitem{MBPS09}
J. Mairal, F. Bach, J. Ponce, and G. Sapiro. Online dictionary learning for sparse
coding. In{\it
Proceedings of the International Conference on Machine Learning
(ICML)}
, 2009a.

\bibitem{MZ93} S. Mallat and Z. Zhang, Matching pursuit with time-frequency
dictionaries, \textit{IEEE Trans. Signal Process}, \textbf{41}, pp.
3397-3415, 1993.

%

\bibitem{SE10}
K. Skretting and K. Engan,
Recursive Least Squares
Dictionary Learning Algorithm, {\it IEEE Transactions on
Signal Processing}, \textbf{58(4)}, pp. 2121-2131, 2010


\bibitem{Shi-13}
 Y. Shi, K. F. Li, Y. L. Yung, H. H. Aumann, Z. Shi, and T. Y. Hou,
A decadal microwave record of tropical air
temperature from AMSU-A/Aqua observations,
{\it Climate Dynamics}, accepted, 2013, DOI 10.1007/s00382-013-1696-x.

%

\bibitem{WH09} Z. Wu and N. E. Huang, Ensemble Empirical Mode Decomposition:
a noise-assisted data analysis method, \textit{Advances in Adaptive
Data Analysis}, \textbf{1}, pp. 1-41, 2009.


\end{thebibliography}
\end{document}